\documentclass[aps,prb,a4paper,twocolumn,superscriptaddress,citeautoscript]{revtex4-1}
\usepackage{color}
\usepackage[english]{babel}
\usepackage{lmodern}
\usepackage[T1]{fontenc}
\usepackage{amsmath}
\usepackage{units}
\usepackage[pdftex]{graphicx}
\usepackage{grffile}
\usepackage{color}
\usepackage[bottom]{footmisc}
\usepackage{booktabs}
\graphicspath{{figs/}}
\newcommand{\LB}{\mathtt{LB}}
\newcommand{\SB}{\mathtt{SB}}
\newcommand{\SSB}{\Sigma_{\SB}}
\newcommand{\SLB}{\Sigma_{\LB}}
\newcommand{\ep}{\varepsilon_{p}}
\newcommand{\esb}{\varepsilon_{\mathrm{SB}}}
\newcommand{\elb}{\varepsilon_{\mathrm{LB}}}
\newcommand{\epsA}{\eps_{\LB}}
\newcommand{\epsB}{\eps_{\SB}}
\newcommand{\er}{\delta}
\newcommand{\tk}{t_{\mathbf{k}}}
\newcommand{\eps}{\varepsilon}
\newcommand{\kv}{\mathbf{k}}
\newcommand{\Deltaeff}{\Delta^{\textrm{eff}}}
\DeclareMathOperator*{\Tr}{Tr}
\begin{document}
\title{Optical spectroscopy and the nature of the insulating state of rare-earth nickelates}
\author{J.~Ruppen}
\affiliation{Department of Quantum Matter Physics, University of Geneva, 24 
Quai Ernest-Ansermet, 1211 Geneva 4, Switzerland}
\author{J.~Teyssier}
\affiliation{Department of Quantum Matter Physics, University of Geneva, 24 
Quai Ernest-Ansermet, 1211 Geneva 4, Switzerland}
\author{O.~E.~Peil}
\affiliation{Department of Quantum Matter Physics, University of Geneva, 24 
Quai Ernest-Ansermet, 1211 Geneva 4, Switzerland}
\affiliation{Coll\`ege de France, 11 place Marcelin Berthelot, 75005 Paris, 
France}
\author{S.~Catalano}
\affiliation{Department of Quantum Matter Physics, University of Geneva, 24 
Quai Ernest-Ansermet, 1211 Geneva 4, Switzerland}
\author{M.~Gibert}
\affiliation{Department of Quantum Matter Physics, University of Geneva, 24 
Quai Ernest-Ansermet, 1211 Geneva 4, Switzerland}
\author{J.~Mravlje}
\affiliation{Jo\v{z}ef Stefan Institute, Jamova~39, Ljubljana, Slovenia}
\author{J.-M.~Triscone} 
\affiliation{Department of Quantum Matter Physics, University of Geneva, 24 
Quai Ernest-Ansermet, 1211 Geneva 4, Switzerland}
\author{A.~Georges}
\affiliation{Department of Quantum Matter Physics, University of Geneva, 24 
Quai Ernest-Ansermet, 1211 Geneva 4, Switzerland}
\affiliation{Coll\`ege de France, 11 place Marcelin Berthelot, 75005 Paris, 
France}
\affiliation{Centre de Physique Th\'eorique, \'Ecole Polytechnique, CNRS, 91128 
Palaiseau Cedex, France}
\author{D.~van der Marel}
\affiliation{Department of Quantum Matter Physics, University of Geneva, 24 
Quai Ernest-Ansermet, 1211 Geneva 4, Switzerland}
\date{\today}
\begin{abstract}
Using a combination of spectroscopic ellipsometry and DC transport measurements, we determine the temperature dependence of the optical conductivity of NdNiO$_3$ and SmNiO$_{3}$ films. 
The optical spectra show the appearance of a characteristic two-peak structure in the near-infrared when the material passes from the metal to the insulator phase. 
Dynamical mean-field theory calculations confirm this two-peak structure, and allow to identify these spectral changes and the associated changes in the electronic structure.
We demonstrate that the insulating phase in these compounds and the associated characteristic two-peak structure are due to the combined effect of bond-disproportionation and Mott physics associated with half of the disproportionated sites. 
We also provide insights into the structure of excited states above the gap.
\end{abstract}
\maketitle
\section{Introduction}
The rare earth nickelates RNiO$_3$ form a remarkable group of materials\cite{Torrance1992,Garcia-Munoz1992,Medarde1997,Catalan2008}.
While LaNiO$_3$ remains metallic down to very low temperatures, all other nickelates undergo a metal-insulator phase transition (MIT) and antiferromagnetic (AF) ordering as the temperature is lowered. 
The two transitions coincide for Pr and Nd but they are distinct, with  $T_{\mathrm{AF}}<T_{\mathrm{MIT}}$, for all rare-earth cations smaller than Nd  (Sm, Gd, and so on down to Lu)~\cite{Medarde1997}.
The mechanism of this MIT, which differs from that of a homogeneous Mott transition, raises questions of fundamental importance.
Furthermore, the possibility of controlling the MIT by chemical substitutions, strain, heterostructures, gating or light pulses\cite{Garcia-Munoz1992,Medarde1997,Catalan2008,Boris2011,Caviglia2012,Scherwitzl2010} make these materials particularly interesting for potential applications. 
For those reasons, nickelates have recently been the subject of intensive research and attention. 

\begin{figure}
\begin{center}
\includegraphics[width=\columnwidth]{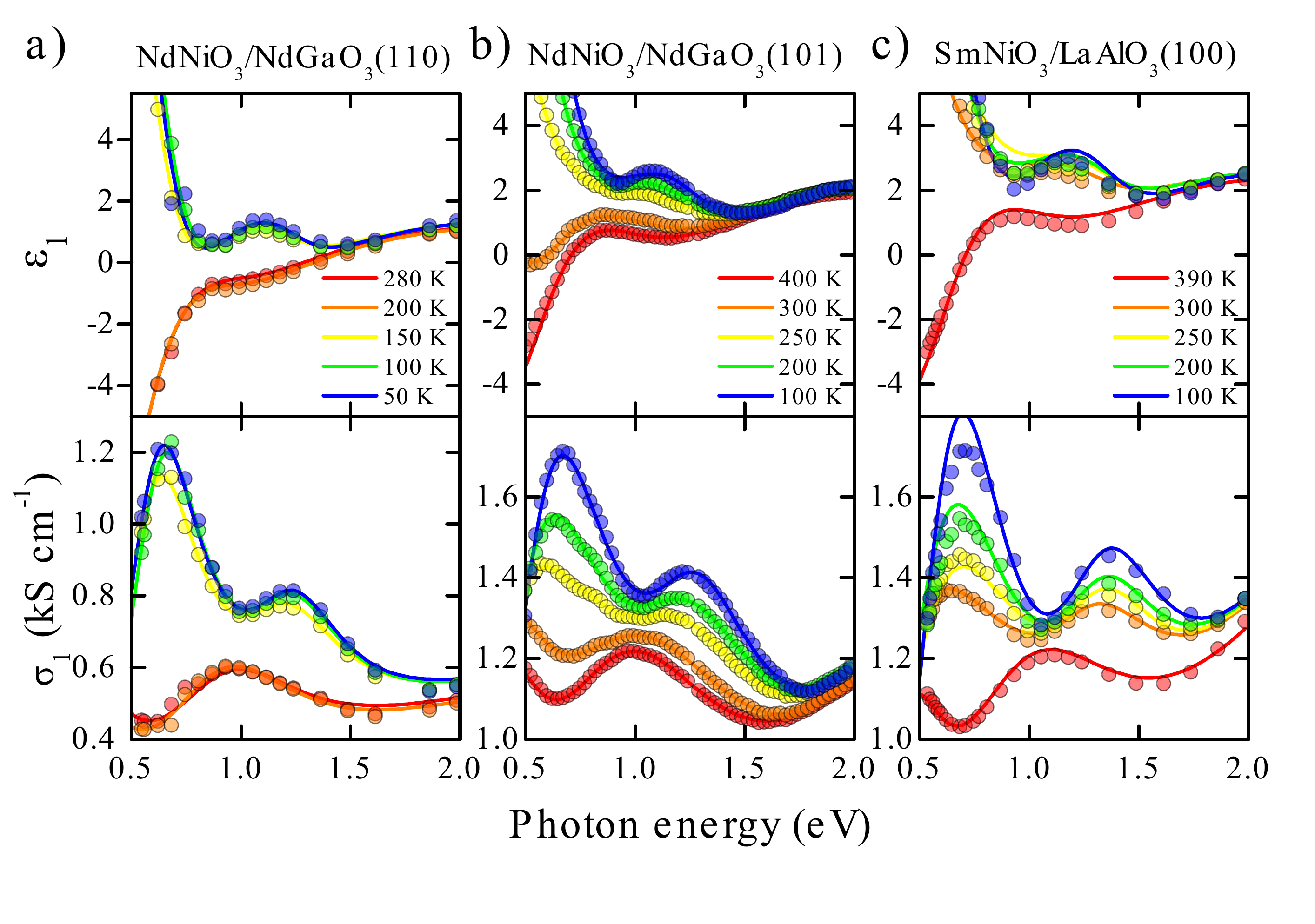}
\caption{\label{fig:RNO_opt_Fig1} Real part of the dielectric function (top) and optical conductivity (bottom) of NdNiO$_3$ on a 
NdGaO$_3$ $(110)$ substrate (left), NdNiO$_3$ on a NdGaO$_3$ $(101)$ substrate (middle) and for 
SmNiO$_3$ on a LaAlO$_3$ $(100)$ substrate (right)}
\end{center}
\end{figure}
\begin{figure*}
\begin{center}
\includegraphics[width=\textwidth]{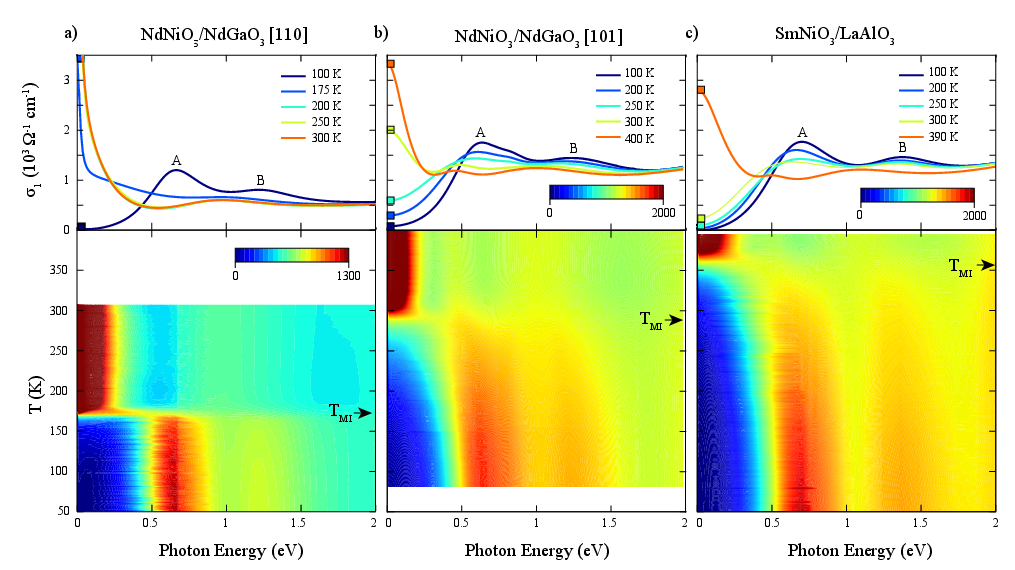}
\caption{\label{fig:RNO_opt_Fig2}
Real part of the optical conductivity for selected temperatures and energy/temperature color maps of samples
a) NNO/NGO-110, 
b) NNO/NGO-101  and 
c) SNO/LAO-001. 
Metal-insulator phase transitions are indicated by arrows on the colormaps. 
$A$ and $B$ designate two peaks in the insulating phase.
Data at 0 eV come from DC measurements.}
\end{center}
\end{figure*}
The insulating phase is characterized by a lowering of the crystal symmetry from orthorombic to monoclinic and by a disproportionation of Ni-O bond lengths:  the NiO$_6$ octahedra undergo a breathing distortion, with alternating long-bond (LB) and short-bond (SB) octahedra on each sublattice. 
This lattice modulation is accompanied by some form of charge-ordering, the precise nature of which has been the subject of debate. 
Early work~\cite{Demourgues1993,Mizokawa2000} emphasized the formation of ligand-holes,  and the importance of the $d^8\underline{L}$ local configuration, in contrast to the $d^7$ configuration corresponding to the nominal Ni$^{3+}$ valence.
This leads to a physical picture for the charge-ordering in which Ni-O bonds are involved (rather than Ni atomic sites). 
In an extreme limit of this picture, LB octahedra are associated with the $d^8$ configuration (with a large local moment) and SB ones with $d^8\underline{L}^2$ (with the Ni local moment screened by the two ligand holes) \cite{Johnston2014,Park2012}. 
Recently, theoretical work has provided support to this physical picture: in Ref.~\onlinecite{Park2012} the MIT was explained as a `site-selective' Mott transition  associated with the $d^8$ LB sites, and in Ref.~\onlinecite{Subedi2015} a corresponding low-energy  description was proposed, focusing on the strongly hybridized Ni-O states with $e_g$ symmetry. 
In this description, consistently with the proposal of Ref.~\onlinecite{Mazin2007}, an effectively attractive interaction between electrons with parallel spins in different orbitals naturally leads to the formation of a bond density-wave. 
Since the bands are quarter filled, the corresponding doubling of the unit cell opens a gap {\em above} the Fermi level, leaving the material metallic at the band-structure level.
The observed insulating state results in fact from the {\em combination} of unit cell doubling and Mott physics (local moments) at the LB sites.
Although a consistent picture of the MIT appears to be emerging, a direct comparison to experiments is still lacking. 
 
In this paper, we report experimental optical spectra on three different nickelate systems. 
These spectra show a common feature: the appearance of two peaks as the MIT is crossed - hence a `universal' feature of the MIT. 
We show that this provides direct insight into the structure of the insulating phase, and that the 2-peak structure results from the bond-disproportionated nature of the low-$T$ phase, with two kinds of nickel sites.
We perform dynamical mean-field theory (DMFT) calculations within the theoretical framework introduced in Ref.~\onlinecite{Subedi2015}, which are found to reproduce quite well the main features of the optical spectra. 
Based on these calculations, we provide a simple analytical understanding of these main features, and of the relative roles of the Peierls and Mott mechanisms in the MIT of nickelates. 

\section{Experiment}
Scanning tunneling microscopy~\cite{Allen2015} and Terahertz time-domain~\cite{Rana2014} spectroscopy experiments have been interpreted as evidence of a charge-density wave formation.
Previous optical studies of NdNiO$_{3}$ films have already shown that in crossing from the correlated metalllic (mass enhancement of order 4 \cite{Ouellette2010,Stewart2012}) to the insulating phase strong peaks appear at  approximately $0.5$~eV and $1.0$~eV \cite{Katsufuji1995,Stewart2011}.
The Drude spectral weight is redistributed up to at least $5$~eV when the system becomes insulating \cite{Katsufuji1995,Stewart2011,Jaramillo2014}, which was interpeted as an indication of Mott physics \cite{Stewart2011} or effects of electron-phonon interaction \cite{Jaramillo2014}.
However, a clear mechanism of the optical response of the insulating phase has thus far been lacking.

The following thin film/substrate combinations were used in the present study:
NdNiO$_3$ on an $(110)$ oriented NdGaO$_3$ substrate (NNO/NGO-110)\footnote{Note that NdGaO$_3$ is orthorhombic, and that we employ the corresponding notation for the crystal planes. 
The (110) and (101) planes of NdGaO$_3$ correspond to $(001)_{pc}$ and $(111)_{pc}$ in pseudo-cubic notation.}, 
NdNiO$_3$ on NdGaO$_3$ $(101)$ (NNO/NGO-101) and SmNiO$_3$ on LaAlO$_3$ $(001)$ (SNO/LAO-001). 
These high quality epitaxial films were prepared as described in Refs. \onlinecite{Catalano2014,Catalano2015}.
The dielectric function was determined in the range from $0.5$ and $2$~eV using ellipsometry at a reflection angle between 65 and 72 degrees with the surface normal. 
Measurements were performed in steps of $1$ K using a special UHV cryostat with a vacuum better than $10^{-9}$~mbar. 
Data of substrate films and substrates were combined to calculate the complex dielectric function, $\epsilon(\omega)=\epsilon_1(\omega)+i4\pi\sigma_1(\omega)/\omega$, using the Fresnel relations (see Appendix \ref{Appendix:AB}). 
The resulting thin film dielectric functions for all three samples are presented in Fig. \ref{fig:RNO_opt_Fig1} for a limited set of temperatures.
\begin{figure}
\includegraphics[width=\linewidth]{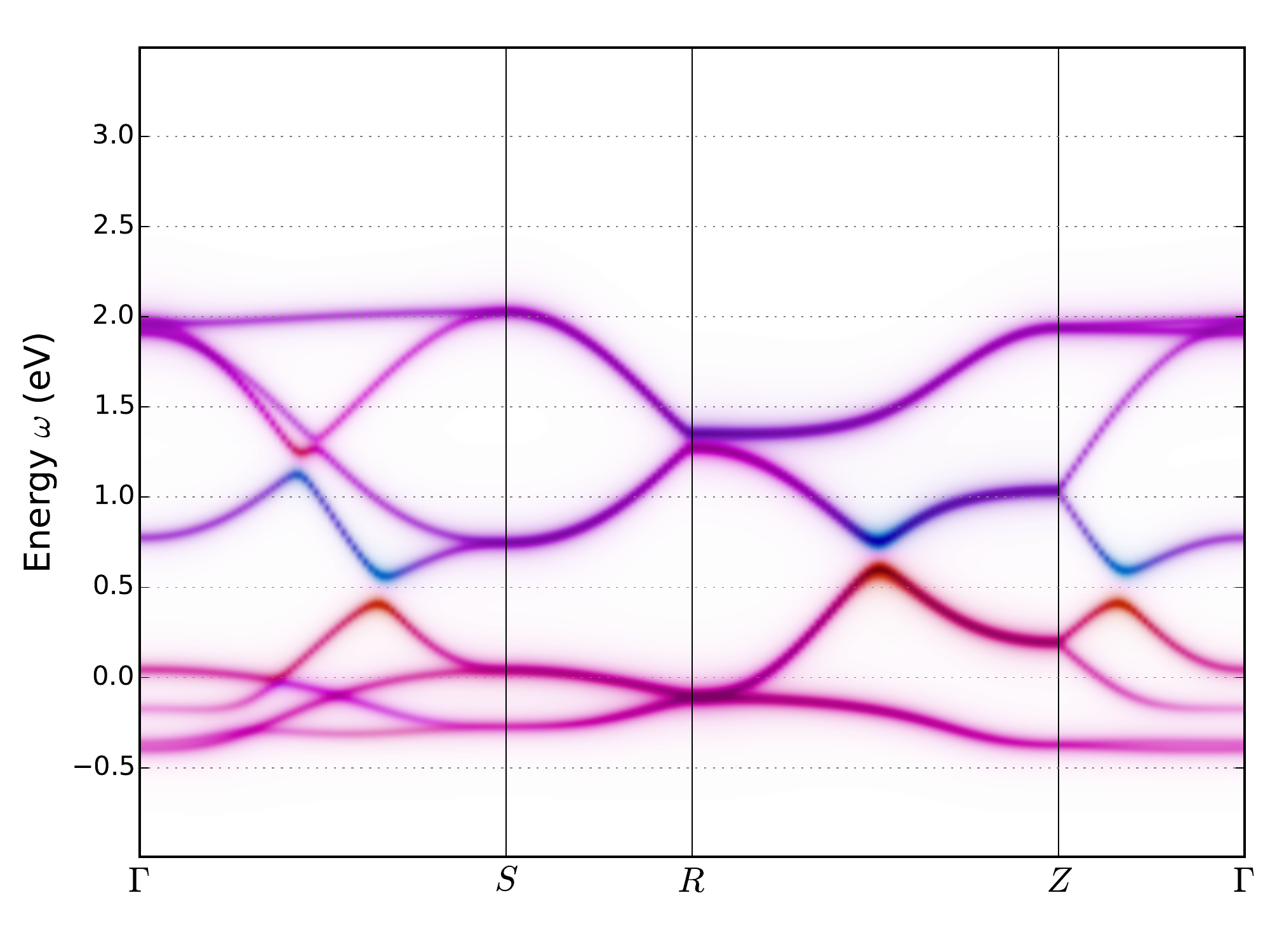}
\caption{\label{fig:RNO_opt_Fig3} The bare(GGA) band structure of the monoclinic phase of SmNiO$_{3}$.
The color represents the site character of the states: LB (red) and SB (blue).
Note the Peierls splitting at an energy +0.5--0.7 eV. The position of the Fermi level is $\varepsilon=0$. }
\end{figure}
DC resistivities of the films were measured as a function of temperature using the four terminal method.
Drude Lorentz-fitting to the DC resistivity (symbols at zero energy in Fig.~\ref{fig:RNO_opt_Fig2}) and the complex dielectric function from $0.5$ to $2$~eV was used to interpolate the optical data below $0.5$~eV. 
While the spectral weight of $\sigma_1(\omega)$ integrated from $0$ to $0.5$~eV is accurately represented by this procedure due to the constraints imposed by simultaneously fitting $\sigma_1(\omega)$ {\em and} $\epsilon_1(\omega)$ \cite{Kuzmenko2007}, fine details such as phonons are not captured in this representation.  

Figure~\ref{fig:RNO_opt_Fig2} shows the energy dependence (upper panels) and energy/temperature color maps (lower panels) of the real part of the optical conductivity for samples NNO/NGO-110 (a), NNO/NGO-101 (b) and SNO/LAO-001 (c).
In the insulating state, at low temperatures, the dominant features of the optical conductivity are two peaks at 0.6 ($A$) and 1.4~eV ($B$) for all three samples (Fig.~\ref{fig:RNO_opt_Fig2} a-c). 
Upon increasing the temperature and passing through the insulator-metal transition, the peaks vanish and a broad 1~eV peak along with a weak feature at 0.5~eV for samples (b) and (c) appear instead. 
Formation of free carriers is clearly visible with the growth of a zero energy mode in the optical conductivity for $\hbar\omega \lesssim 1$~eV (Fig.~\ref{fig:RNO_opt_Fig2}) and a sign change in the real part of the dielectric function (Fig.~\ref{fig:RNO_opt_Fig1}). 

From the metallic to the insulating state, all three samples present a comparable amount of spectral weight, of approximately $3$~eV$^2$, which is transferred from the region below 0.5~eV (representative of the free carriers in the system) to higher energy range which extends to at least $5$~eV pointing to strong correations in the insulating state \cite{Stewart2011}.
\section{{Theoretical calculations}}
To understand the nature of the optical excitations observed experimentally, we have performed DMFT calculations for the bulk low-$T$ phase of SmNiO$_{3}$ (space group $P2_{1}/n$), within the low-energy framework introduced in Ref.~\onlinecite{Subedi2015} (calculation details in Appendix \ref{Appendix:C}). 
This approach involves only the states with $e_g$ symmetry resulting from the anti-bonding combinations  of Ni-$3d$ and O-$2p$ states. 
At the band-structure level, this corresponds to eight bands, reflecting the four Ni-sites per unit cell with  two $e_g$ states per site, and a total occupancy of $4$ electrons per unit cell (one per site on average). 
The bare (GGA) band structure of monoclinic SmNiO$_3$ is displayed in Fig.~\ref{fig:RNO_opt_Fig3}.
The bands with  $e_g$ character form a well-isolated set of bands of total bandwidth  $\simeq 2.3$~eV, separated by a gap of $\sim0.5 eV$  from the low-lying $t_{2g}$ and oxygen states (not shown on Fig.~\ref{fig:RNO_opt_Fig3}). 
At the LDA/GGA level, these materials are metallic in both the orthorhombic and monoclinic  structure, with the Fermi level crossing the $e_g$ manifold. 
As clearly seen on Fig.~\ref{fig:RNO_opt_Fig3}, the breathing distortion (bond disproportionation)  in the monoclinic structure leads to the opening of a Peierls-like gap in the energy-range  $0.5-0.7$~eV. 
This gap separates 4 lower-lying bands with dominantly LB   character and four higher-lying bands with dominantly SB character.
Due do the breathing distortion of the low-$T$ phase the local on-site energies of LB and SB sites are split by $\Delta_{s}\simeq 0.25$~eV. 
This in turn results in the opening  of a Peierls-like gap in the band-structure (of magnitude $\sim\Delta_s$)  at an energy of order $+0.5$~eV above Fermi level corresponding to half-filling (two electrons per site).
This Peierls mechanism alone is therefore insufficient to account for the insulating nature of this phase, and  correlations play an essential role. 

\begin{figure}
\begin{center}
\includegraphics[width=\columnwidth]{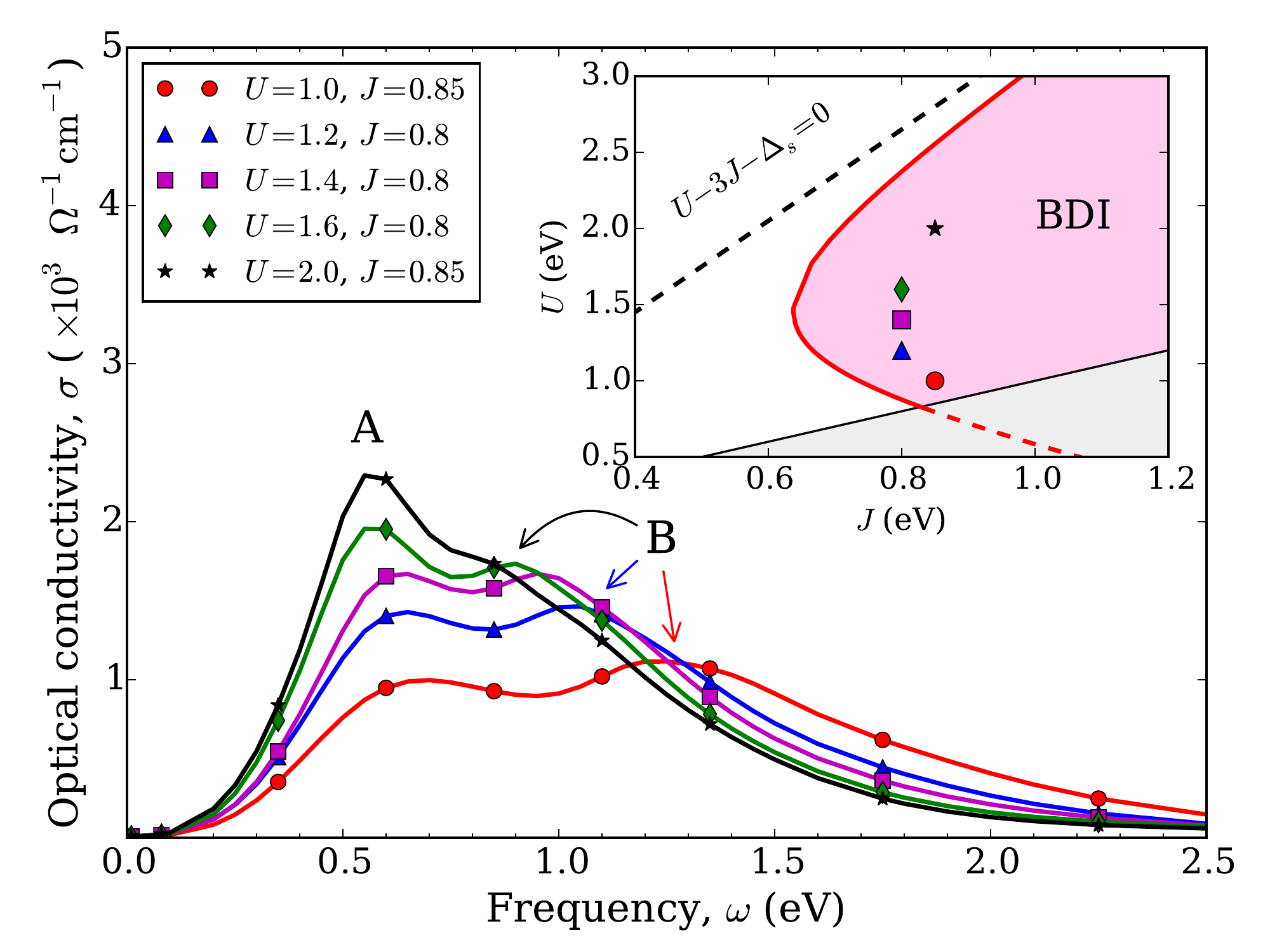}
\caption{\label{fig:RNO_opt_Fig4}
Calculated optical spectra for several values of $U$ and $J$. 
The two peaks are denoted by $A$ (constant position) and $B$ (varying position).
Inset: Phase boundary of the bond-disproportionated insulating state.  
The symbols indicate the values of $U,J$, chosen such that the leading edge is kept approximately constant of order $0.5$~eV.}
\end{center}
\end{figure}

As was demonstrated in Ref.~\onlinecite{Subedi2015}, considering the Coulomb repulsion $U$ and Hund's coupling $J$ acting within the set of $e_{g}$ states allows one to describe the MIT provided that $U - 3J \lesssim \Delta_{s}$. 
As illustrated on the partial phase diagram in the inset of Fig.~\ref{fig:RNO_opt_Fig4},  a `bond-disproportionated insulator' (BDI) phase is found in this regime, in which the $e_g$ occupancy is modulated, with a smaller value on the SB sites and  a larger one on the LB sites. 
Orbital polarization is weak in this BDI state, with both  $e_g$ orbitals approximately equally occupied on each site.

We applied an \textit{ab initio} LDA+DMFT approach to the low-$T$ phase of SmNiO$_{3}$ and calculated the electronic spectral functions and the optical conductivity using the Kubo formalism (see Appendix \ref{Appendix:C}).
We display in Fig.~\ref{fig:RNO_opt_Fig4} the calculated optical spectra,  for a set of values of $U$ and $J$ within the BDI phase. 
The values, indicated in the inset of Fig.~\ref{fig:RNO_opt_Fig4}, are chosen in such a way that the leading edge is roughly constant and close to the observed experimental value ($\sim 0.5$~eV). 
In agreement with experimental data, all the theoretical spectra demonstrate the presence of two peaks (denoted by $A$ and $B$ in the figure).
While the position of peak $A$ is fixed by the choice of parameters, both the position of peak $B$ and its relative intensity increase as one moves from the upper to the lower boundary of the phase diagram, {\em i.e.} as $U-3J$ becomes more negative and the disproportionation increases.   

\begin{figure}
\begin{center}
\includegraphics[width=\columnwidth]{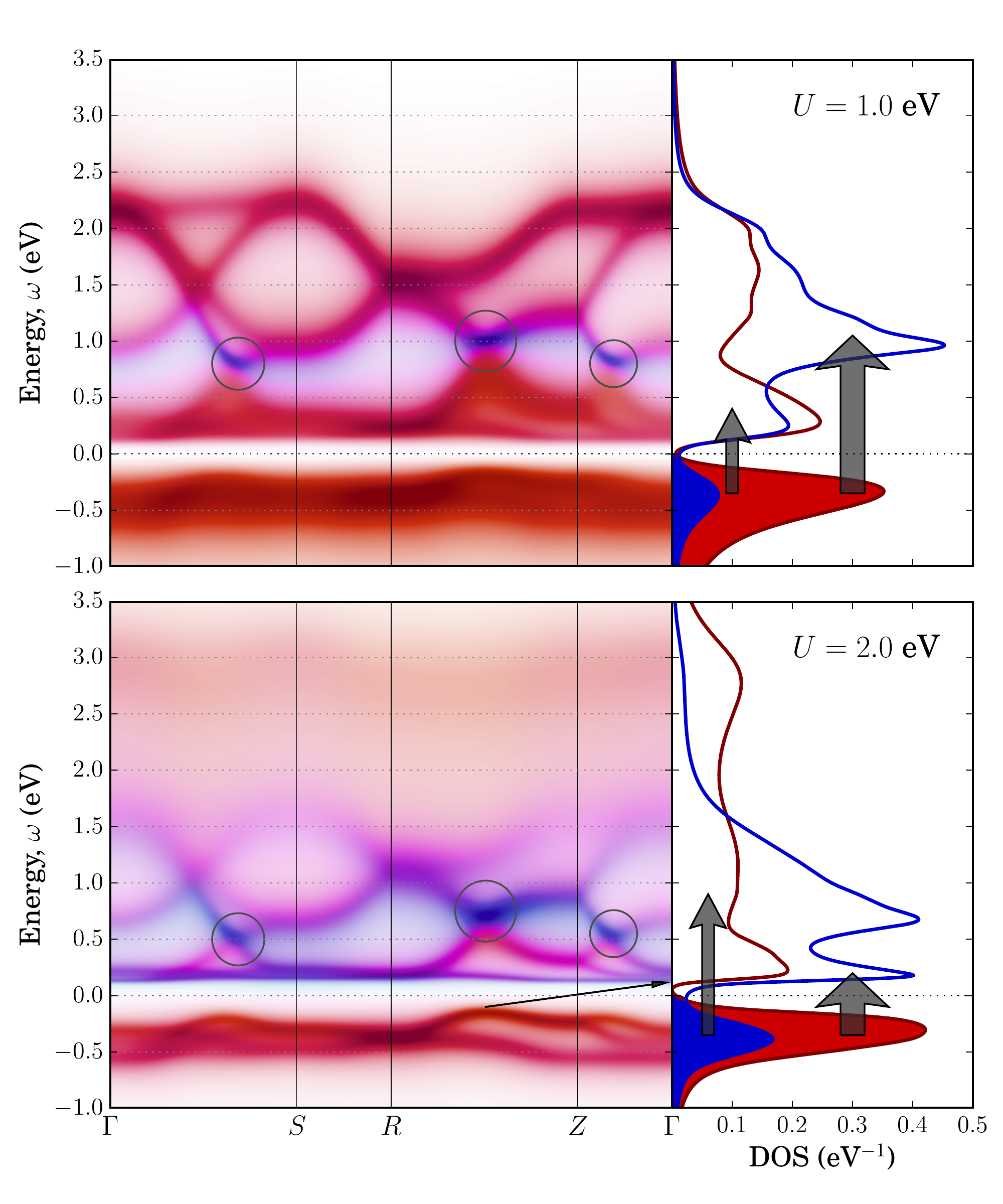}
\caption{\label{fig:RNO_opt_Fig5} 
Momentum-resolved spectral function (color intensity map) for two selected parameter sets: $U = 1.0$~eV, $J = 0.85$~eV (top) and $U =2.0$ eV, $J = 0.85$~eV (bottom).   
The colors represent the site character of a state: red for LB, blue for SB  (violet for a mixed LB/SB character).
A darker (lighter) tone corresponds to higher (lower) spectral intensity.
The circles indicate the energy-momentum locations where the Peierls pseudo-gap opens.  
The indirect Mott-like gap is indicated by the black arrow connecting the highest occupied states between $R$ and $Z$ and the lowest unoccupied states at $\Gamma$-point.
Side-panel: momentum-integrated spectral functions (density of states) for 
LB (red) and SB (blue) sites, with arrows indicating the optical transitions corresponding to the two peaks (see text).}
\end{center}
\end{figure}

To identify the optical transitions associated with these two peaks, we display in Fig.~\ref{fig:RNO_opt_Fig5} the momentum-resolved spectral functions plotted for the two extreme points (both at $J=0.85$~eV):  $U = 2.0$~eV (smaller disproportionation) and $U = 1.0$~eV (larger disproportionation). 
Three sets of states can be identified, split by a correlation-induced indirect gap at the Fermi level and by a pseudo-gap (originating from the Peierls LB/SB site modulation) at around $+0.5$~eV.   
The Peierls pseudo-gap is clearly seen in the density of states (side panels of Fig.~\ref{fig:RNO_opt_Fig5}) and the momentum locations at which it opens are indicated by circles in the main panel.  
The states below Fermi level always have dominant LB character, in accordance with the largest occupancy of LB states. 
In contrast, the nature of the lowest unoccupied band immediately above the gap changes from dominantly LB at $U=1$~eV to dominantly SB at $U=2$~eV.

The optical transitions responsible for the lower-energy peak $A$ are the ones across the insulating gap, while the second peak is due to optical transitions across the Peierls pseudo-gap, as indicated by arrows on Fig.~\ref{fig:RNO_opt_Fig5}. 
The current operator has only inter-site matrix elements, with largest nearest-neighbor components coupling sites with different characters. 
This explains why the first peak has higher relative intensity when the states on either side of the gap have different characters, {\em i.e.} on the upper side of the BDI phase boundary (smaller disproportionation). 
\begin{figure}
\begin{center}
\includegraphics[width=\columnwidth]{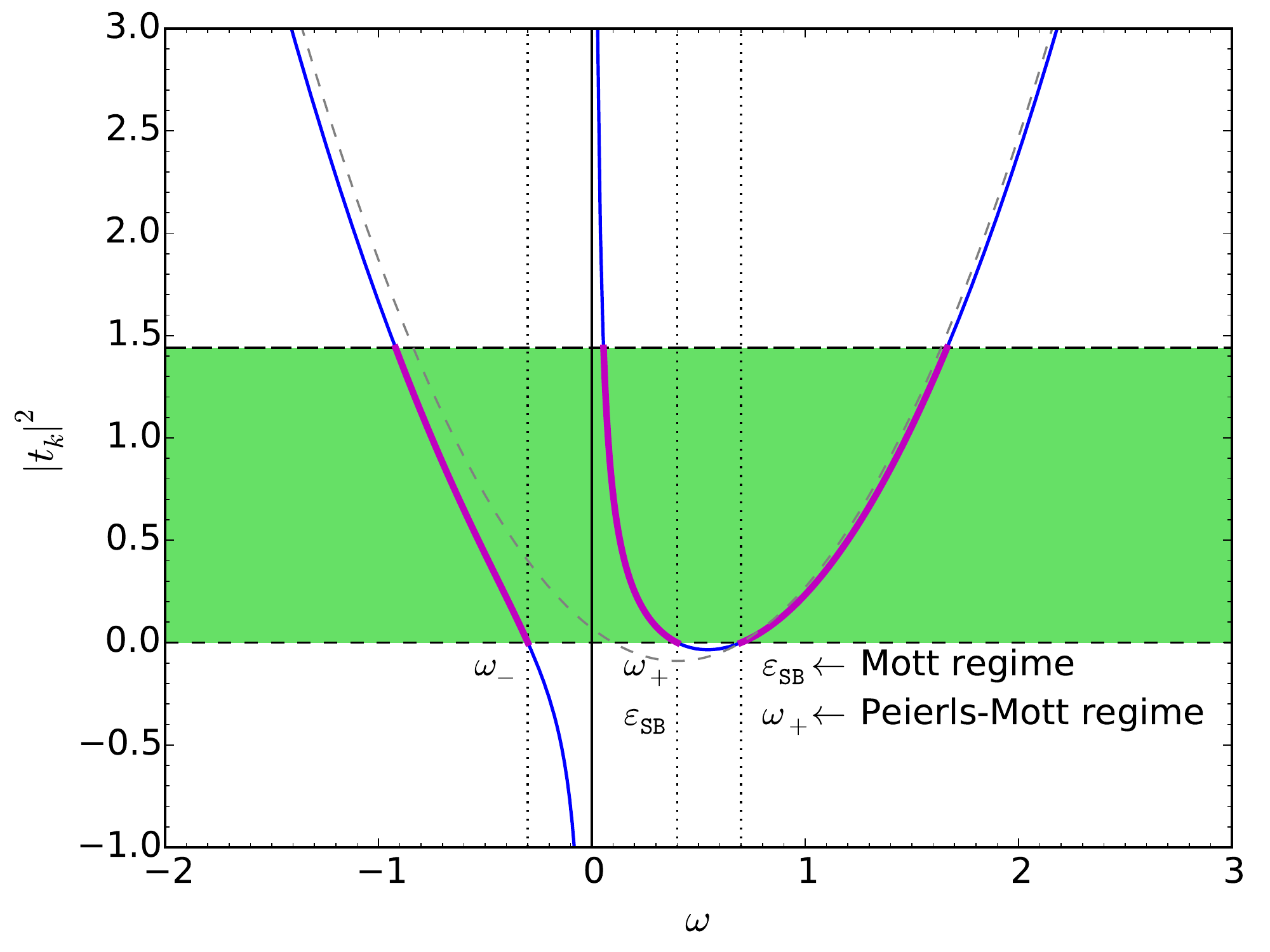}
\caption{\label{fig:RNO_opt_Fig6}Left-hand side of the quasi-particle equation Eq.~\eqref{eq:qp_eq} and graphical construction of the three QP branches. 
The green (shaded) area shows the region of allowed values of the RHS: $0 \le |t_{k}|^{2} \le W^{2} / 4$, where $W$ is the bandwidth.
$\omega_{i}$, $i = +,-$, and $\omega = \epsB$ are the roots of the equation for $t_{k} = 0$. 
The order of the roots $\omega_{+}$ and $\epsB$ depends on the regime (Mott or Mott-Peierls, see text).}
\end{center}
\end{figure}

\section{{Analysis and discussion}}
\begin{figure}
\includegraphics[width=\linewidth]{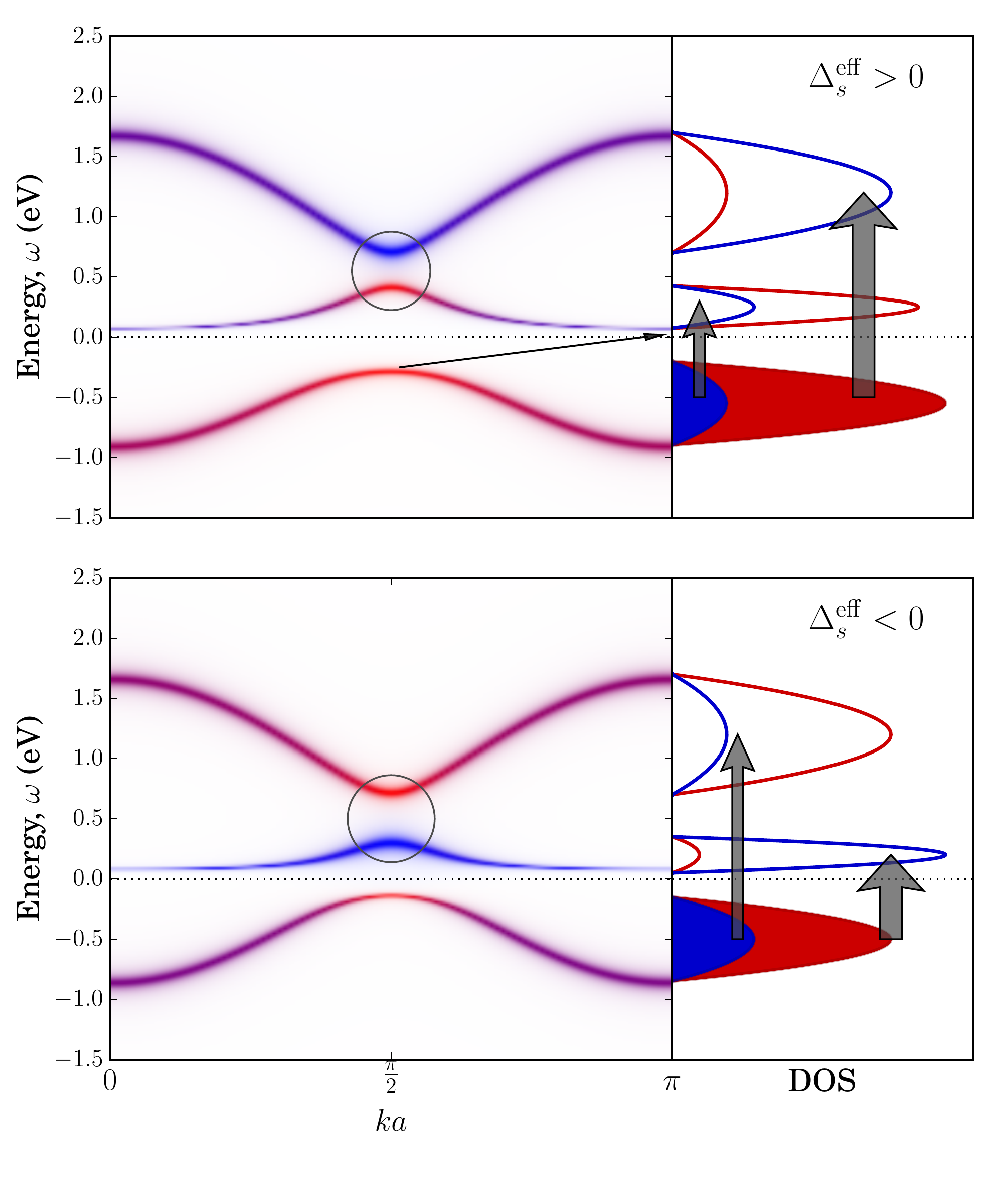}
\caption{\label{fig:RNO_opt_Fig7}Quasiparticle band structure, density of states and main optical transitions of the simple model discussed in the text. 
Top: Mott regime, bottom: Mott-Peierls regime. 
Colors indicate the LB/SB character, as above. 
The indirect Mott-like gap is indicated with the thin arrow in the main panel. }
\end{figure}
Fig.~\ref{fig:RNO_opt_Fig5} reveals that the dominant site character of the lowest unoccupied states above the gap is different for the two values of $U$. 
Specifically, it is LB-like for $U = 1.0$ (corresponding to the lower part of the BDI region in the phase diagram) and SB-like for $U = 2.0$ (upper part).
To understand better the electronic structure and, in particular, the structure of unoccupied states we first note that, as further detailed in Appendix \ref{Appendix:D}, the self-energies in the BDI phase can be well described at low energy by 
\begin{align}
\SSB^\prime(\omega)-\mu = & \esb &
\SLB^\prime(\omega)-\mu = & \frac{\er^2}{\omega-\ep}\,+\,\elb
\label{eq:simple_self}
\end{align}
These expressions have a simple physical meaning. 
The lower occupancy SB sites are weakly correlated and hence have an approximately constant self-energy. 
The LB sites, in contrast, have a  self-energy typical of a Mott insulator, with a pole-like divergence at $\omega=\ep\simeq 0$ which is responsible for the opening of the insulating gap (with a magnitude controlled by the energy scale $\delta$).
This is consistent with the `site-selective Mott' picture of Ref.~\onlinecite{Park2012}. 

Simplifying further, let us consider a model with only two sites (LB and SB) per unit cell and nearest-neighbor hopping $\tk$, so that the non-interacting hamilltonian reads:  
\begin{align}
H^0_{\kv} =
\begin{bmatrix}
\epsA^{(0)}  &  t_{\kv} \\
t^{*}_{\kv} & \epsB^{(0)}
\end{bmatrix},
\end{align}
The dispersion, $\omega=\omega_{\mathbf{k}}$ of quasiparticles (QP) is then determined 
from the zeros of the determinant of $\omega+\mu-H^0_{\kv}-\hat{\Sigma}(\omega)$, 
leading to ($\ep$ is neglected below):
\begin{align}\label{eq:qp_eq}
\left(\omega - \elb - {\er^{2}}/{\omega}\right) (\omega - \esb) = \left| \tk \right|^{2},
\end{align}
where 
$\eps_{\LB,\SB} = \eps_{\LB,\SB}^{(0)} - \mu + \Sigma_{\LB,\SB}^{\infty}$.
This cubic equation has three QP branches, which are displayed on Fig.~\ref{fig:RNO_opt_Fig7} for a simple one-dimensional tight-binding band $\tk=W(1+e^{i2ka})/4$ 
(with $W$ the bandwidth). 

To analyze this equation we plot the LHS as a function of $\omega$, as depicted in Fig.~\ref{fig:RNO_opt_Fig6}, bearing in mind that the allowed states are limited by the range of values of the RHS, $0 \leq |t_{k}|^{2} \leq W^{2}/4$.
(The plot is done for $\epsB > 0$, since the BDI state has site occupancies $n_{\LB} > n_{\SB}$.) 
From the figure we immediately see that the Mott gap at around zero frequency is an indirect one, as expected, and the Peierls gap is direct.
For $k=\pm \pi/2a\equiv k_P$, which is the Fermi momentum of the half-filled system at which the Peierls gap opens, one has $\tk=0$ and the three roots read:  
\begin{align}
\omega_{-} = & \frac{1}{2}\left(\epsA - \sqrt{\epsA^{2} + 4 \er^{2}}\right), \\
\omega_{+} = & \frac{1}{2}\left(\epsA + \sqrt{\epsA^{2} + 4 \er^{2}}\right), \\
\omega = & \epsB.
\end{align}
The occupied QP states correspond to $\omega_{-}<0$ and have predominantly LB character. 
The insulating gap is always indirect, corresponding to transitions between the top of the occupied band at $k=k_P$ 
and the bottom of the lowest unoccupied band at $k=0$ ($\Gamma$-point). 
It can be estimated as: 
\begin{equation}
\Delta_g \simeq \frac{4\delta^2\epsB}{W^2} + \frac{1}{2}\left(\sqrt{\epsA^{2} + 4 \er^{2}}-\epsA\right)
\label{eq:gap_text}
\end{equation}
Note that it vanishes for $\er=0$ as expected.

The nature of the lowest unoccupied branch above the insulating gap depends on the sign of: 
\begin{equation}
\Deltaeff_{s}=\epsB-\omega_{+}=\epsB- \frac{1}{2}\left(\epsA + \sqrt{\epsA^{2} + 4 \er^{2}}\right)
\end{equation}
whose magnitude $|\Deltaeff_s|$ is the Peierls direct gap renormalized by correlations, 
which separates the two unoccupied branches and opens at $k = k_P$ (as indicated by circles in Fig.~\ref{fig:RNO_opt_Fig7}).  
For $\Deltaeff_{s} > 0$ the lowest branch of unoccupied states has dominantly LB character: this corresponds to the regime of large disproportionation in which the almost half-filled LB band undergoes a Mott transition (top panels of Fig.~\ref{fig:RNO_opt_Fig5} and Fig.~\ref{fig:RNO_opt_Fig7}, corresponding to the lower boundary of the BDI phase). 
For $\Deltaeff_{s} < 0$ the situation is reversed, and the states above the insulating gap are dominantly SB (bottom panels in Figs.~\ref{fig:RNO_opt_Fig5}, \ref{fig:RNO_opt_Fig7}; corresponding to the upper boundary of the BDI phase with smaller disproportionation). 
In this `Mott-Peierls' regime, the Mott mechanism has pushed the upper Hubbard band above the unoccupied band of SB states, and the states on either sides of the insulating gap have different characters (analogously to what happens in a charge-transfer insulator). 
Relative intensity and separation of the two peaks in the experimental data, suggest that the nickelates studied here  may be more in the Mott-Peierls regime or in the crossover between the two regimes.

\section{Conclusions}
In summary, using ellipsometry we have measured the detailed temperature dependence of the optical conductivity spectra of strained RNiO$_3$ epitaxial thin films. 
The insulator is characterized by the occurrence of a conspicuous double-peak structure. 
\textit{Ab initio} calculations and model considerations indicate that this optical signature reveals the peculiar structure of the insulating state suggested earlier \cite{Johnston2014,Park2012,Subedi2015}.
Specifically, the two peaks in the optical conductivity of the insulating phase can be assigned to transitions from the lower Hubbard band to unoccupied bands split by a renormalized Peierls gap.
Moreover, the model reveals two possible regimes with the lowest unoccupied states being of either LB or SB character, with the considered nickelate systems being close to the crossover between the two regimes. 
This provides another possibility of tailoring the properties of these materials by controlling the charge carrier density via stoichiometry or heterostructure engineering.

\begin{acknowledgements}
We gratefully acknowledge discussions with G.~A.~Sawatzky and A.~Subedi, whom we also thank for previous collaboration.
This project was supported by the Swiss National Science Foundation (project 200021-146586 and NCCR MARVEL). 
Computer time was provided by CSCS under project s497. 
The research leading to these results has received funding from the European Research Council under the European Union's Seventh Framework Programme (FP7/2007-2013) / ERC Grant Agreement Nr. 319286 (Q-MAC).
\end{acknowledgements}
%
%
%
\appendix
\section{Ellipsometry of thin films\label{Appendix:AB}}
\begin{table}[]
\begin{ruledtabular}
\begin{tabular}{l l l l l}
 &  & \multicolumn{2}{c}{$T_{MI}/T_{N}$ (K)} \\
Film / Thickness / Substrate & Strain  & Film  & Bulk\\
\hline
NdNiO$_3$/ 30 nm / NdGa$O_3$ $(110)$ & +1.1\% & 175/175  &  200/200 \\
NdNiO$_3$/ 17 nm / NdGa$O_3$ $(101)$ & +1.1\% & 300/225  &  200/200 \\
SmNiO$_3$/ 10 nm / LaAl$O_3$ $(001)$ & $-0.1\%$ & 360/205  &  400/200   
\end{tabular}
\end{ruledtabular}
\caption{\label{table:1} Summary of the three film/substrate systems studied. 
Film thickness is small enough to ensure monocrystallinity and homogeneous strain. The properties in column "Bulk" are from Ref. \onlinecite{lacorre1991}.}
\end{table}
In this paper we study the properties of epitaxial thin films on different substrates. In table \ref{table:1} selected properties of these films are summarized and compared to properties of the  corresponding bulk materials.

\begin{figure}
\begin{center}
\includegraphics[width=\columnwidth]{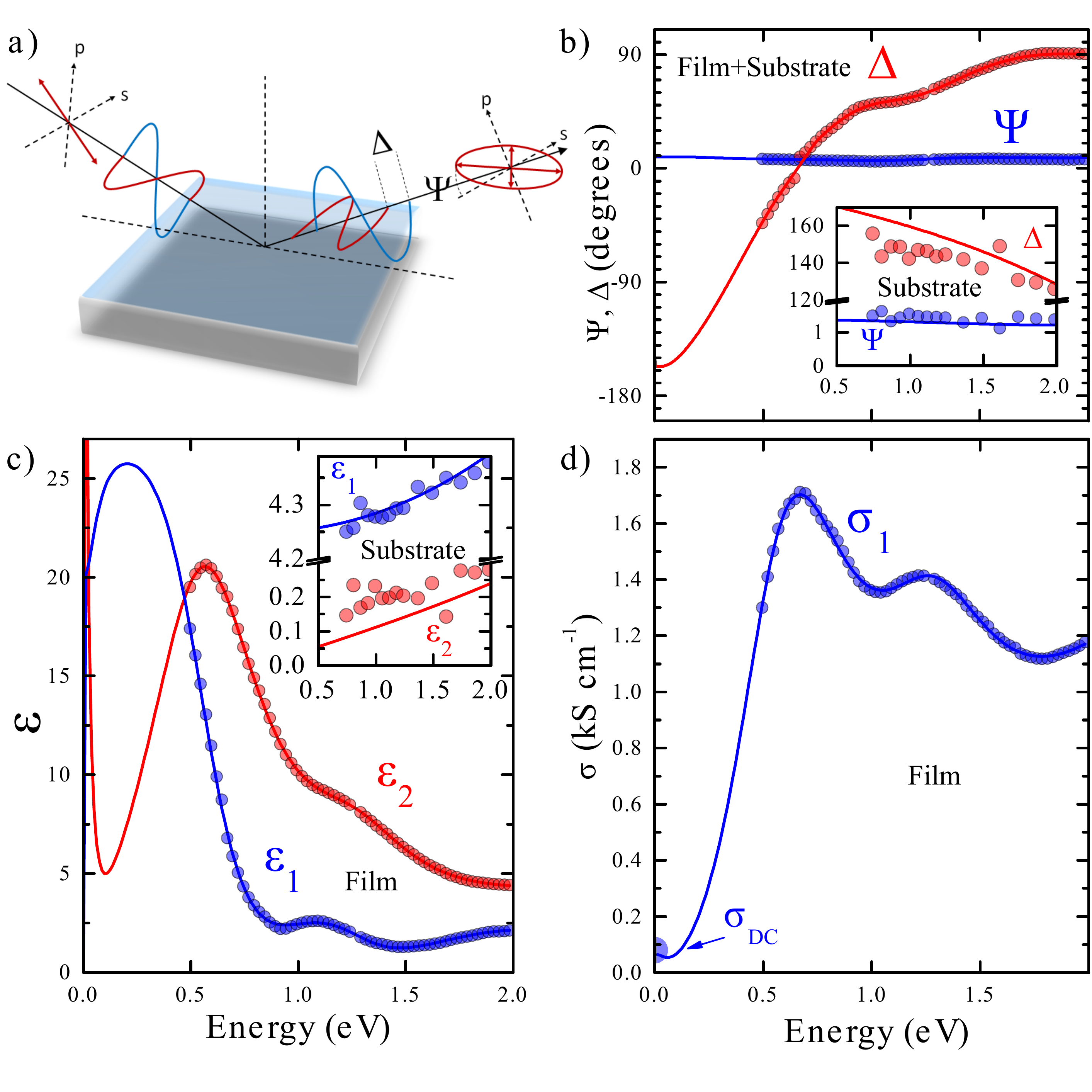}
\caption{\label{fig:RNO_opt_Fig8} (a) Sketch of the optical experiment determining the ellipsometric parameters $\Delta$ (phase shift between p- and s-polarized component of the light) and $\Psi$ (argument of the ratio of p- and s-polarized amplitude). 
(b) $\Psi$ and $\Delta$ spectra of 17 nm thick NdNiO$_3$ on NdGaO$_3$(101) at 100 K measured with an angle of incidence of 69 degrees, shown with the Drude-Lorentz fit.  
(c) Thin film dielectric function calculated from panel (b) with the method explained in the main text. 
The insets of (b) and (c)  show the same parameters for the NdGaO$_3$(101) substrate. 
(d) Optical conductivity corresponding to $\epsilon_2(\omega)$ in panel (c), with the DC conductivity. }
\end{center}
\end{figure}

\begin{figure}
\begin{center}
\includegraphics[width=\columnwidth]{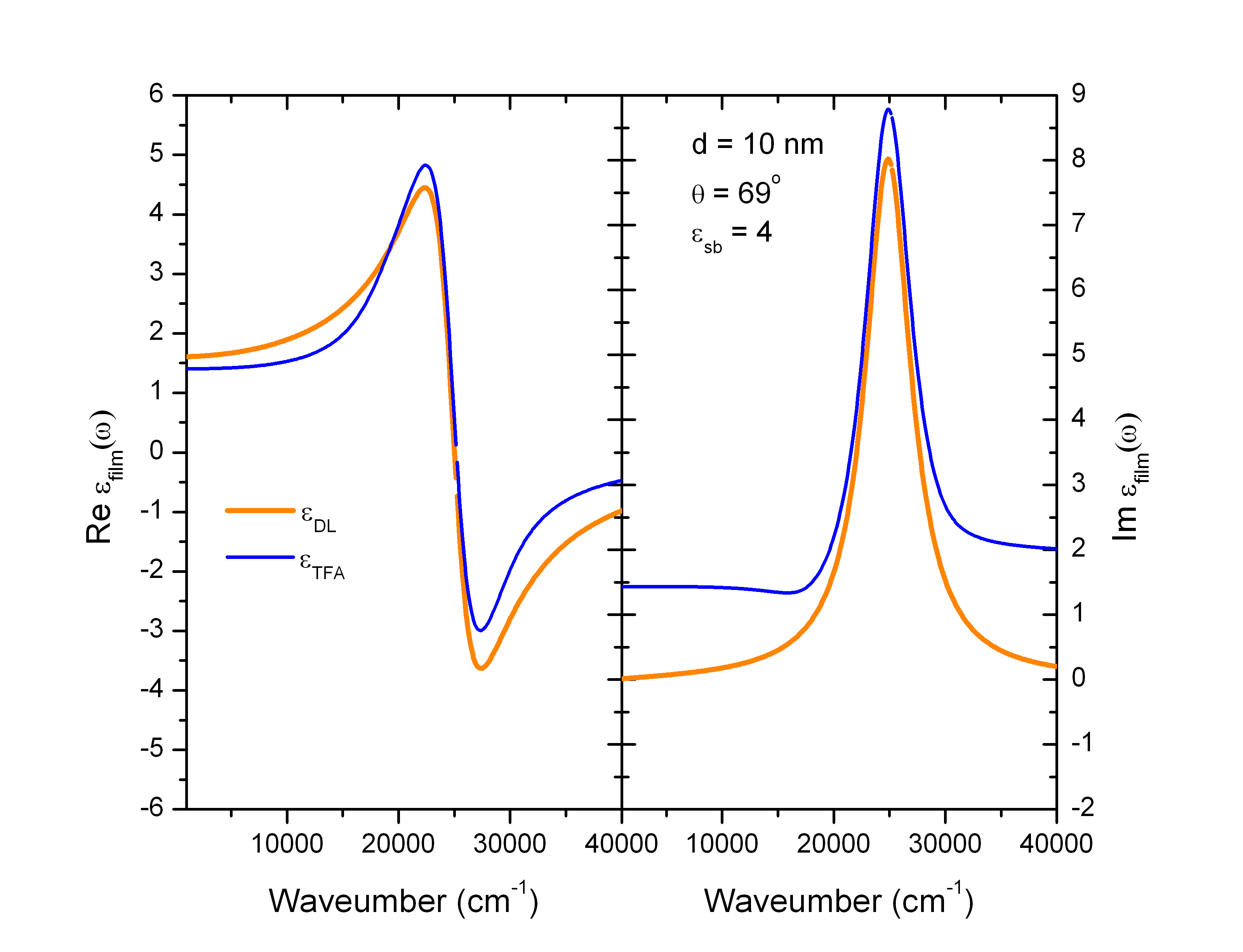}
\caption{\label{fig:RNO_opt_Fig9} Output obtained when applying the thin-film approximation, Eq. \eqref{fresnelthin}, to ellipsometric data $r_p/r_s$ of a film of finite thickness on a substrate ($\epsilon_{TFA}$, blue curves). 
A Drude-Lorentz parametrization ($\epsilon_{DL}$, orange curves) was used to generate the ellipsometric $r_p/r_s$. 
Parameters for the substrate dielectric function, film thickness, and angle of incidence are indicated in the legend.}
\end{center}
\end{figure}

The complex ratio of p-polarized over s-polarized reflectivity is given by $\rho=r_p/r_s=\tan\Psi e^{i\Delta}$. 
The coefficient $\Psi$ and the absolute value of the phase difference of p- and s-polarized light, $|\Delta|$, are the key parameters determined in spectroscopic ellipsometry experiment (Fig.\ref{fig:RNO_opt_Fig8}). 
For of a film with dielectric constant $\epsilon_f$, on a substrate with dielectric constant $\epsilon_s$, the ellipsometric coefficients follow from the relation
\begin{align}\label{fresnelfull}
&\rho=\frac{1-\alpha\cos\theta}{1+\alpha\cos\theta}
\frac{\cos\theta+\beta}{\cos\theta-\beta}
\\
&\mbox{with}\nonumber\\
&\alpha=\frac{\epsilon_f}{\eta_f}
\frac{\epsilon_s\eta_f-i\epsilon_f\eta_s\tan\phi}
{\epsilon_f\eta_s-i\epsilon_s\eta_f\tan\phi}\hspace{2mm};\hspace{2mm}
\beta=\eta_f\frac{\eta_s-i\eta_f\tan\phi}
{\eta_f-i\eta_s\tan\phi}\nonumber\\
&\eta_{s}=\sqrt{\epsilon_{s}-\sin^2\theta}\hspace{2mm};\hspace{2mm}
\eta_{f}=\sqrt{\epsilon_{f}-\sin^2\theta}\hspace{2mm};\hspace{2mm}
\phi=\frac{\omega d}{c}\eta_{f}\nonumber
\end{align}
where $\theta$ is the angle of incidence with the surface normal, $d$ is the film thickness, $\omega$ the angular frequency and $c$ the speed of light. 
To obtain the thin film dielectric function $\epsilon_f$, analytic inversion of Eq. \eqref{fresnelfull} is not a practical approach, in particular since the non-linearity of this expression gives rise to multiple solutions. 
Instead we used a two-step approach. 
We begin by expanding the above expression in leading order of $\phi$, corresponding to the limit of an ultrathin film. 
In this limit analytic inversion is straightforward:
\begin{align}\label{fresnelthin}
\epsilon_{TFA}&=\left(\frac{1+\epsilon_s+A}{2}\right)\pm\sqrt{\left(\frac{1+\epsilon_s+A_s}{2}\right)^2-\epsilon_s}
\nonumber\\
&\mbox{where}\nonumber\\
A_s&=
\frac{c}{\omega d}
\frac
{
(\rho/\rho_0-1)(1-\epsilon_s)(\epsilon_s\cos^2\theta-\sin^2\theta)
}
{
2i\cos\theta(\epsilon_s-\sin^2\theta)
}
\end{align}
and $\rho_0$ is the $r_p/r_s$ ratio of the bare substrate. 
One of these two ($\pm$) solutions has a negative imaginary part, and should be discarded for this reason. 
In Fig. \ref{fig:RNO_opt_Fig9} we give an example of the output for $\epsilon_f(\omega)$ following this approach, when starting from a known thin-film dielectric function generated by a Drude-Lorentz multi-oscillator model. 
The difference between the input dielectric function, and the result obtained using the thin-film approximation is a consequence of the fact that Eq. \eqref{fresnelthin} is strictly valid in the limit $\phi\rightarrow 0$. 
The dielectric function corresponding to the full solution of Eq. \eqref{fresnelfull}  is obtained by inserting a multi-oscillator Drude-Lorentz model for $\epsilon_f(\omega)$ 
\begin{align}
\epsilon_f(\omega)&=\epsilon_{\infty}+\sum_j \frac{\omega_{p,j}^2}
{\omega_j^2-\omega(\omega+i\gamma_j)}
\end{align}
and adjusting the Drude-Lorentz parameters to fit the expression for $\rho$ to the experimental data. 
The output for  $\epsilon_f(\omega)$ obtained by the latter method follows rather closely the dielectric function obtained from the thin-approximation, Eq. \eqref{fresnelthin}. 
In the main body of the paper we will present the dielectric function obtained with the latter method. 
For all data discussed we used the first-mentioned method for the purpose of a sanity check, as illustrated by Fig. \ref{fig:RNO_opt_Fig9}.

While $\Delta>0$ for isotropic bulk materials, for a substrate/thin film system $\Delta$ can become negative for certain frequencies. 
For all our spectra the sign of  $\Delta$ was unambiguously fixed by the Kramers-Kronig constraints on the real- and imaginary part of the corresponding dielectric function. 
In particular we obtained that  in the insulating phase $\Delta<0$ for $\hbar\omega \lesssim 0.5$ eV. 
\begin{table}[ht]
\begin{ruledtabular}
\begin{tabular}{c| c c c| c c c }
& & 100 K  &&& 300 K & \\
Oscillator & ${\omega_p}_j$ & $\omega_j$  & $\gamma_j$ & ${\omega_p}_j$ & $\omega_j$  & $\gamma_j$  
     \\
\hline
NdGaO$_3(110)$ &$\epsilon_{\infty}=1.0$&&&$\epsilon_{\infty}=1.0$&   &  \\
1 & 132481 & 73498 & 22516 & 130863 & 72379 & 23688  \\ 
\hline
NdGaO$_3(101)$ &$\epsilon_{\infty}=1.0$&&&$\epsilon_{\infty}=1.0$&   &  \\
1 & 132481 & 73498 & 22516 & 130863 & 72379 & 23688  \\ 
\hline
LaAlO$_3(001)$ &$\epsilon_{\infty}=2.9$&&&$\epsilon_{\infty}=2.9$&  &  \\
1 &12269 & 24894 & 32465 & 12344 & 26245 & 34905  \\
2 &49871 & 50225 & 11552 & 51511 & 50756 & 12633  \\
\end{tabular}
\end{ruledtabular}
\caption{\label{table:2} Drude-Lorentz parameters of the dielectric function of NdGaO$_3(110)$,  NdGaO$_3(101)$ and LaAlO$_3(001)$ at two different temperatures. 
The notation for the crystal planes of NdGaO$_3$ correspond to the orthorhombic structure, {\em i.e.} (110) and (101) correspond to $(001)_{pc}$ and $(111)_{pc}$ if we approximate the actual structure by the pseudo-cubic symmetry.}
\end{table}

An additional point for $\sigma(\omega)$ was obtained at $\omega=0$ by measuring the DC resistivity of the films. 
Simultaneous fitting of $\sigma(0)$ and the $\Psi(\omega)$ and $\Delta(\omega)$ spectra between 0.5 and 4 eV to a Drude-Lorentz-model allowed us to interpolate the spectra between 0 and 0.5 eV. 
The complete set of temperature dependent spectra of $\epsilon_1(\omega)$ and $\sigma_1(\omega)$ are shown in Fig. \ref{fig:RNO_opt_Fig1}. 
\begin{table}[h!]
\begin{ruledtabular}
\begin{tabular}{c |c c c |c c c }
 & & 100 K  &&& 300 K & \\
 Oscillator & ${\omega_p}_j$ & $\omega_j$  & $\gamma_j$ & ${\omega_p}_j$ & $\omega_j$  & $\gamma_j$ 
      \\
\hline
NdNiO$_3$ &$\epsilon_{\infty}=1.96$& &&$\epsilon_{\infty}=1.92$&   &  \\
 1 & 857    &    0  & 346    & 13840  &    0  & 1607      \\
 2 & 15721  & 5284  & 3742   & 7688   &     0 & 229      \\
 3 & 10257  & 10083 & 4728   & 13161  & 7683  & 7326      \\
 4 & 33302  & 23402 & 34504  & 34516  & 24539 & 38743      \\
\end{tabular}
\end{ruledtabular}
\caption{\label{table:3} Drude-Lorentz parameters of the dielectric function of NdNiO$_3$ 
on NdGaO$_3$ $(110)$ at two different temperatures. }
\end{table}
\section{General Theoretical Methodology\label{Appendix:C}}
Theoretical optical spectra of SmNiO$_{3}$ are obtained numerically using an
\textit{ab initio} description within the DFT+DMFT framework.
First, the crystal structure is obtained by full structure relaxation using the Vienna \textit{ab-initio} simulation package (\textsc{vasp}) \cite{paw_vasp,vasp1,vasp2}.
To this end, we have employed the generalized gradient approximation (GGA) plus Hubbard $U$ (GGA + $U$) with the fully rotationally invariant interaction term  \cite{Liechtenstein1995} and parameters $U = 5.0$ eV and $J = 1.0$ eV.
The plane-wave cut-off is chosen to be $E_{\textrm{cut}} = 550$ eV and  $k$-mesh consists of $7\times7\times5$ points.
The structure relaxation by means of a conjugate-gradient algorithm starting from an orthorhombic 20-atom unit cell resulted in the monoclinic phase (space group $P2_{1}/n$) exhibiting a breathing  structure distortion characterized by long-bond (LB) and short-bond (SB) octahedra, similar to what was found experimentally in small-cation nickelates,  such as LuNiO$_{3}$.

The low-energy model considered in Ref.~\onlinecite{Subedi2015} and in the present paper is  obtained by constructing Wannier functions corresponding to the full  $e_g$ manifold, {\em i.e.} to the energy window  $[-0.6, 2.6]$~eV.   
This is done using projected localized orbitals (PLO) \cite{Amadon2008}. 
The resulting Ni-centered Wannier functions are delocalized, having substantial weight  on oxygen ions \cite{Subedi2015}, due to the strong hybridization between Ni and O characteristic of these compounds with small or negative charge-transfer energy. 
The breathing distortion and Peierls gap leads to a difference of on-site energies $\Delta_{s}$ in the low-energy $e_g$ hamiltonian, which in the case of SmNiO$_{3}$ is of order $\Delta_{s} \simeq 0.25$ eV (practically the same value as that for LuNiO$_{3}$).    

Local electronic correlations are described by means of a Hubbard-Kanamori Hamiltonian formulated in terms of low-energy $e_{g}$ states (see details in Ref.~\onlinecite{Subedi2015}) with only density-density interactions.
The model is solved within the DFT+DMFT approach, as implemented within the Wien2TRIQS framework\cite{wien2k,TRIQS,Aichhorn2009}.  
The \textsc{TRIQS} library~\cite{TRIQS} implementation of the hybridization-expansion continuous-time Monte Carlo (CT-HYB) solver~\cite{Gull2011} was used. 
The optical conductivity $\sigma(\omega)$ is evaluated using the Kubo formula with neglected vertex corrections, i.e.:
\begin{align}
\sigma_{xx}(\omega) = & \frac{2\pi}{\omega} \int_{-\infty}^{\infty} d\eps\,
[f(\eps) - f(\eps+\omega)]\times \\ 
& \times\frac{1}{V}\sum_{\kv} \Tr\left\{J^{x}(\kv)A(\kv,\eps)J^{x}(\kv)A(\kv,\eps+\omega)
\right\},\nonumber
\end{align}
where $J^{x,y,z}(\kv)\equiv J^{x,y,z}_{\mu\nu}(\kv)$ are current matrix elements, $A(\kv, \eps) \equiv A_{\mu\nu}(\kv, \eps)$ momentum resolved spectral function, and the trace is over orbital (band) indices $\mu$, $\nu$.

The key point is that the parameter range relevant to the physics of nickelates lies in the region $U - 3J \lesssim \Delta_s$. 
The phase diagram for this region was mapped out for LuNiO$_{3}$ in Ref.~\onlinecite{Subedi2015} and has now been recalculated for SmNiO$_{3}$. 
The main features turned out to be the same as for LuNiO$_{3}$. 
Specifically, for $0.7 \mathrm{ eV } \lesssim J \lesssim 1.1$ eV the monoclinic phase is a so-called bond-disproportionated insulator (BDI), while the orthorhombic phase remains metallic.

\section{Frequency-dependence and analytical continuation of the self-energies\label{Appendix:D}}
\begin{figure}
\includegraphics[width=\linewidth]{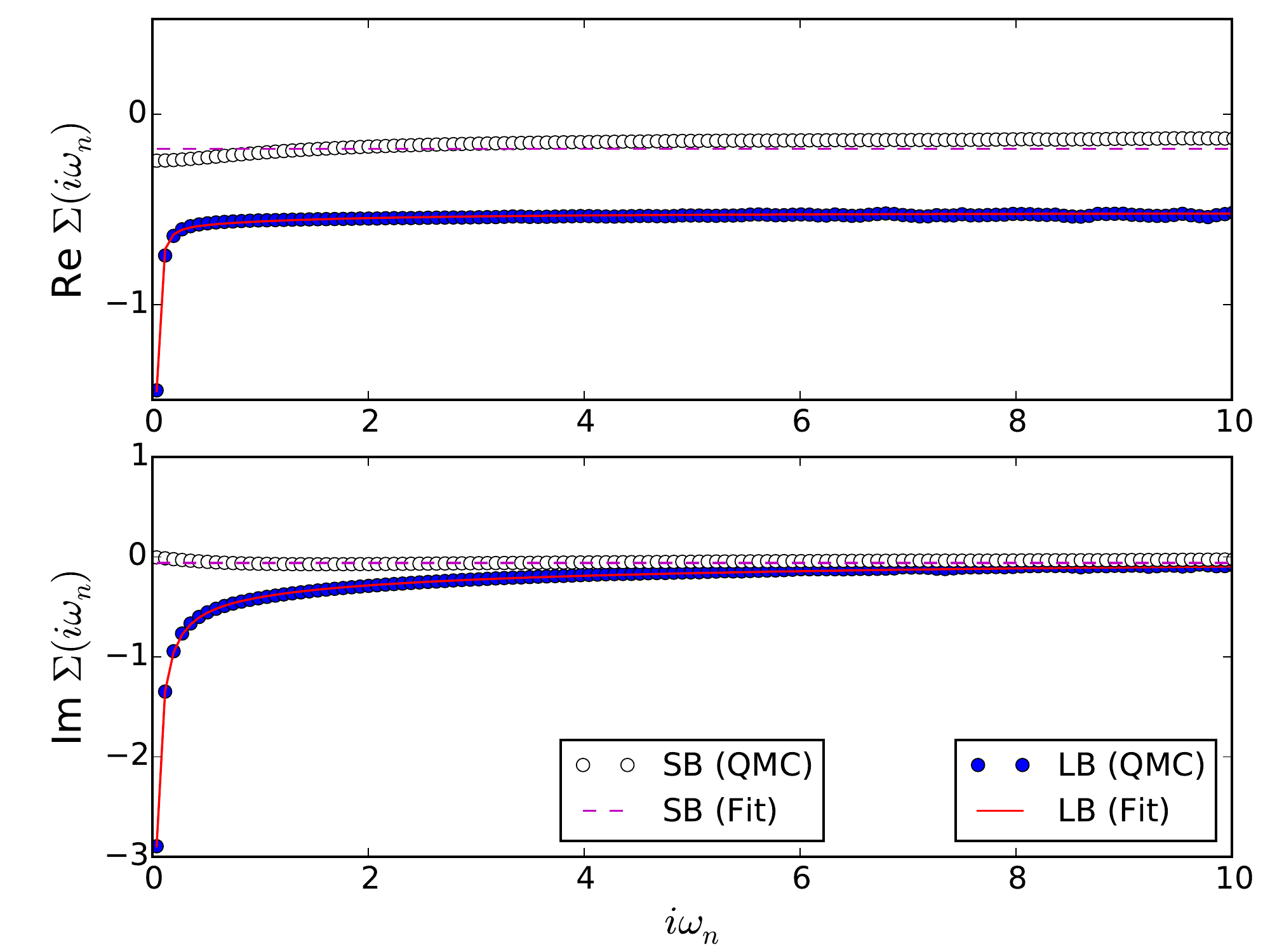}
\includegraphics[width=\linewidth]{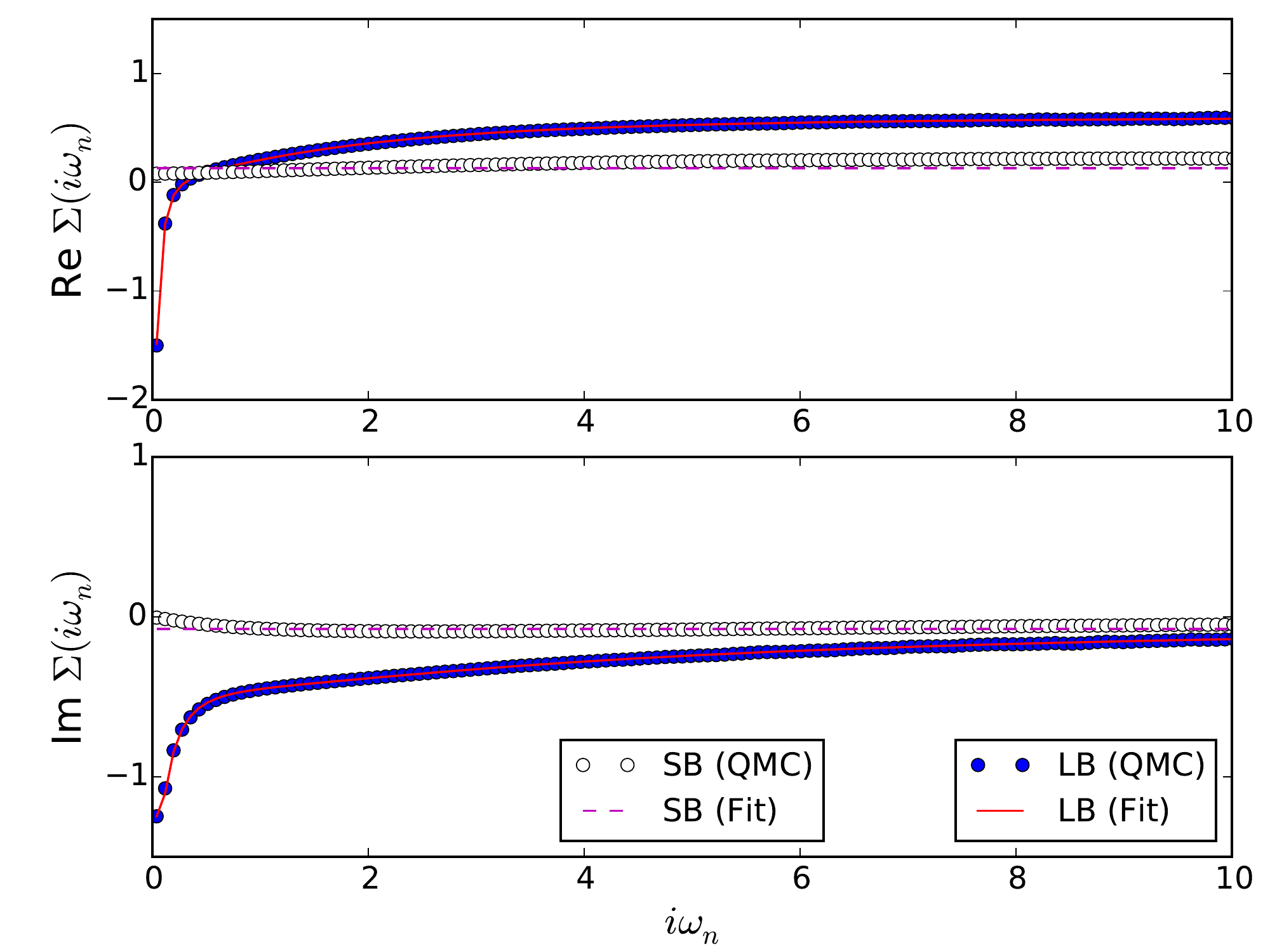}
\caption{\label{fig:RNO_opt_Fig10} Examples of the analytical fit (LB: red solid, SB: magenta dashed) of the QMC self-energy (LB: blue filled circles, SB: green empty circles).
The LB self-energy is fit with a two-pole ansatz given by Eq.~\eqref{eq:fit2pole}, while the SB self-energy is fit with a constant.
Top two panels: $U = 1.0$ eV; bottom two panels: $U = 2.0$ ($J = 0.85$ eV in both cases).}
\end{figure}

Once the DFT+DMFT self-consistency cycle converges, the optical conductivity can be obtained from the interacting Green function (or self-energy) by means of the Kubo-Greenwood formalism, with vertex correction neglected.
This requires the analytical continuation of the DMFT self-energy on the real axis. 
To this end, we have first inspected the self-energies at Matsubara frequencies (see the data for the LB self-energy in Fig.~\ref{fig:RNO_opt_Fig10}).
The self-energies at LB and SB sites behave very differently.
The LB self-energy exhibits a strongly singular behavior at small frequencies suggesting that a representation by a rational fraction \cite{Savrasov2005} is adequate. 
In fact, we have found that the following two-pole ansatz is sufficient to represent the LB self-energy with high accuracy:
\begin{align}
\Sigma_{\LB}(\omega) = \Sigma^{\infty}_{\LB} +
\frac{p_{1}}{\omega - p_{2} - i p_{3}} +
\frac{p_{4}}{\omega - p_{5} - i p_{6}}.
\label{eq:fit2pole}
\end{align}

\begin{figure}
\includegraphics[width=\linewidth]{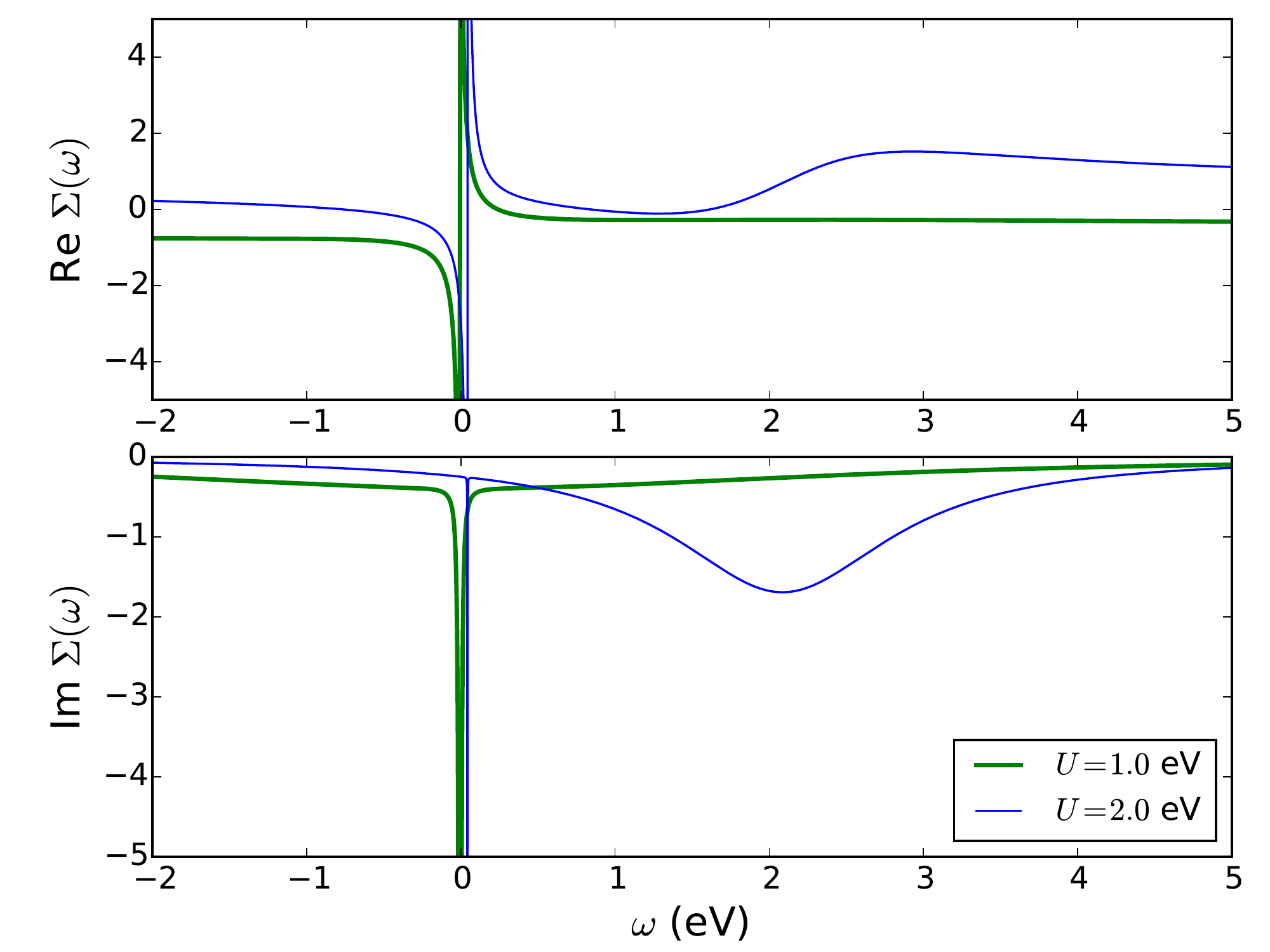}
\caption{\label{fig:RNO_opt_Fig11} Analytical fit of the LB self-energy on the real axis.
Green: $U = 1.0$, blue: $U = 2.0$ ($J = 0.85$ eV in both cases).
The LB self-energy is fitted with a two-pole ansatz given by Eq.~\eqref{eq:fit2pole}, with parameters presented in Table~\ref{table:4}.}
\end{figure}

The fitting parameters $p_{i}$ (as well as $\Sigma^{\infty}_{\LB}$) have been obtained by least-square optimization, with the parameters being bounded to physically meaningful ranges of values (ensuring the analytical behavior of the self-energy). 
We have checked that this ansatz provides a high-quality and reliable fit for all values of $U$ and $J$ corresponding to the BDI phase. 
Examples of the fit for the two limiting cases $U = 1.0$ eV and $U = 2.0$ eV (both for $J = 0.85$ eV) are demonstrated in Fig.~\ref{fig:RNO_opt_Fig10}, where the fits are compared to the actual QMC data at Matsubara frequencies. 
Fig.~\ref{fig:RNO_opt_Fig11} shows the real-frequency LB self-energy and the corresponding fit parameters are presented in Table~\ref{table:4}.

\begin{table}[!h]\centering
\begin{tabular*}{\linewidth}{@{\extracolsep{\stretch{1}}}crrrrrrr@{}}\toprule
$U$ (eV) & $\Sigma^{\infty}_{\LB}$ & $p_{1}$ & $p_{2}$ & $p_{3}$ & $p_{4}$ & $p_{5}$ & $p_{6}$ \\
\midrule
$1.0$ & -0.518 &  0.120 & -0.015 &  0.003 &  1.070 & 0.564 & 2.690 \\
$2.0$ &  0.608 &  0.127 & -0.057 &  0.000 &  1.411 & 2.137 & 0.749 \\
\bottomrule
\end{tabular*}
\caption{Parameters of the two-pole fit to the LB self-energy used in Fig.~\ref{fig:RNO_opt_Fig10} (for $J = 0.85$). 
All parameters are in eV, except for $p_1,p_4$ (eV$^2$). }
\label{table:4}
\end{table}

As to the SB self-energy, it has essentially a featureless structure with a flat real part and very small imaginary part (Fig.~\ref{fig:RNO_opt_Fig10}), which has been approximated by a complex constant (using the least-square fit over a range of frequencies spanning $\sim 4$ eV).

This analysis of the structure of the self-energies has inspired  the description of the electronic structure of the BDI phase presented below.

\section{General considerations and simple model\label{Appendix:E}}
We present here details of the simple model analysis introduced in the main text. 
One of the key experimental observations is that the optical spectrum in the insulating phase consists of two distinct peak (around 0.6 and 1.4 eV). 
As has already been pointed out in Ref.~\onlinecite{Stewart2011} these features correspond to transitions within the $e_{g}$ manifold, and hence their origin can be addressed in the framework of our low-energy $e_g$ description. 

We recall that the bandstructure of the monocliic phase (Fig.~\ref{fig:RNO_opt_Fig3}) has a Peierls  gap separating two manifolds of states, corresponding to the disproportionation between LB (lower manifold) and SB sites (higher manifold).
We also emphasize that optical (current) matrix elements are inter-site and hence dominantly couple LB and SB sites. 

In the limit of very large disproportionation, the transition into the BDI insulator can be considered as a Mott transition on the sublattice of LB sites \cite{Park2012,Subedi2015}.
In the band picture this corresponds to opening of a Mott gap in the lower manifold (LB) of $e_g$ bands. 
This results in three bands of quasiparticle states, one below the Mott gap and two above,  and the two observed optical transitions can be interpreted as transitions between these three set of states. 

This can be captured on a qualitative level by considering a simple model with just two sites per unit-cell (LB and SB) and nearest-neighbour hopping between them,  described by the band Hamiltonian: 
\begin{align}
H^0_{\kv} =
\begin{bmatrix}
\epsA^{(0)}  &  t_{\kv} \\
t^{*}_{\kv} & \epsB^{(0)}
\end{bmatrix}
\end{align}
where $\epsA^{(0)},\epsB^{(0)}$ correspond to bare LB and SB site energies, respectively, $t_{\kv}$ is the hopping amplitude between the sites.
To get the desired site splitting $\Delta_{s} = \epsB^{(0)} - \epsA^{(0)}$ one can choose the parameters to be $\epsA^{(0)} = -\Delta_{s}/2$, $\epsB^{(0)} = +\Delta_{s}/2$.

Inspired by the analysis of the self-energies in Appendix \ref{Appendix:D}, we can adopt a very simple ansatz for the self-energies on both types of sites. 
The SB self-energy can be treated as a constant and for the LB self-energy we can consider a single-pole representation appropriate at low-energy, namely: 
\begin{align}
\Sigma_{\LB}(\omega) = & \frac{\er^{2}}{\omega} + \Sigma_{\LB}^{\infty}, \\
\Sigma_{\SB}(\omega) = & \Sigma_{\SB}^{\infty},
\end{align}
with $\er$ (essentially equal to $\sqrt{p_1}$ above) giving the effective interaction strength and the constants $\Sigma^{\infty}_{\LB,\SB}$ renormalizing the on-site energies. 
These constants depend both on the interaction strength and site occupancies. 
In this expression, we have neglected the fact that the pole of the LB self-energy is slightly offset from zero-energy. 
Indeed, the parameter $p_2$ is always small - it is smallest, as expected from low-energy particle-hole symmetry, when the LB band is almost half-filled (strong disproportionation)-. 

The $k$-resolved spectral functions of this model read:
\begin{align}
A(\kv,\omega) = & -\frac{1}{\pi} \mathrm{Im}
\begin{bmatrix}
\omega - \epsA - \frac{\er^{2}}{\omega} + i0^{+} &  t_{\kv} \\
t^{*}_{\kv} & \omega - \epsB + i0^{+}
\end{bmatrix}^{-1},
\label{eq:akw}
\end{align}
where we have introduced effective on-site energies
$\eps_{\LB,\SB} = \eps_{\LB,\SB}^{(0)} - \mu + \Sigma_{\LB,\SB}^{\infty}$.
The corresponding equation for the eigenvalues is Eq. \ref{eq:qp_eq} of the main text, which yields the dispersion of quasiparticles (QP)  excitations, is thus a cubic equation which has three roots at each $k$-point, resulting in three QP bands. 
  
The characters of the bands can be determined from the corresponding eigenvectors at every $k$-point.
However, to understand the qualitative behavior it is sufficient to examine the $k$-points for which $t_{k} = 0$. 
In this case the matrix in Eq.~\eqref{eq:akw} becomes diagonal and the bands, thus, possess pure site characters.
Moreover, the QP equation is readily factorized when the RHS is zero, which gives the following roots:
\begin{align}
\omega_{-} = & \frac{1}{2}\left(\epsA - \sqrt{\epsA^{2} + 4 \er^{2}}\right), \\
\omega_{+} = & \frac{1}{2}\left(\epsA + \sqrt{\epsA^{2} + 4 \er^{2}}\right), \\
\omega = & \epsB,
\end{align}
where the first two roots correspond to LB bands and the third to a SB band.

The magnitude of the renormalized Peierls gap is therefore given by $|\Deltaeff_s|$, with:   
\begin{equation}
\Deltaeff_s = \epsB-\omega_{+}=\epsB- \frac{1}{2}\left(\epsA + \sqrt{\epsA^{2} + 4 \er^{2}}\right)
\end{equation}
which can be interpreted as the effective site-energy difference between SB and LB sites, renormalized by interactions. 

The indirect Mott gap can be estimated by observing (Fig.~\ref{fig:RNO_opt_Fig6}) that it involves a transition between the occupied (valence) band at $t_k=0$ and the lowest unoccupied (conduction) band at $k=0$ ($|t_k|=W/2$). 
At that point, the QP energy of the conduction band is close to zero-energy. 
This can be used in Eq.~\ref{eq:qp_eq} to obtain the following estimate for the insulating gap: $\Delta_g \simeq 4\delta^2\epsB/W^2-\omega_{-}$, hence: 
\begin{equation}
\Delta_g \simeq \frac{4\delta^2\epsB}{W^2} + \frac{1}{2}\left(\sqrt{\epsA^{2} + 4 \er^{2}}-\epsA\right)
\label{eq:gap}
\end{equation}
\begin{figure}
\includegraphics[width=\linewidth]{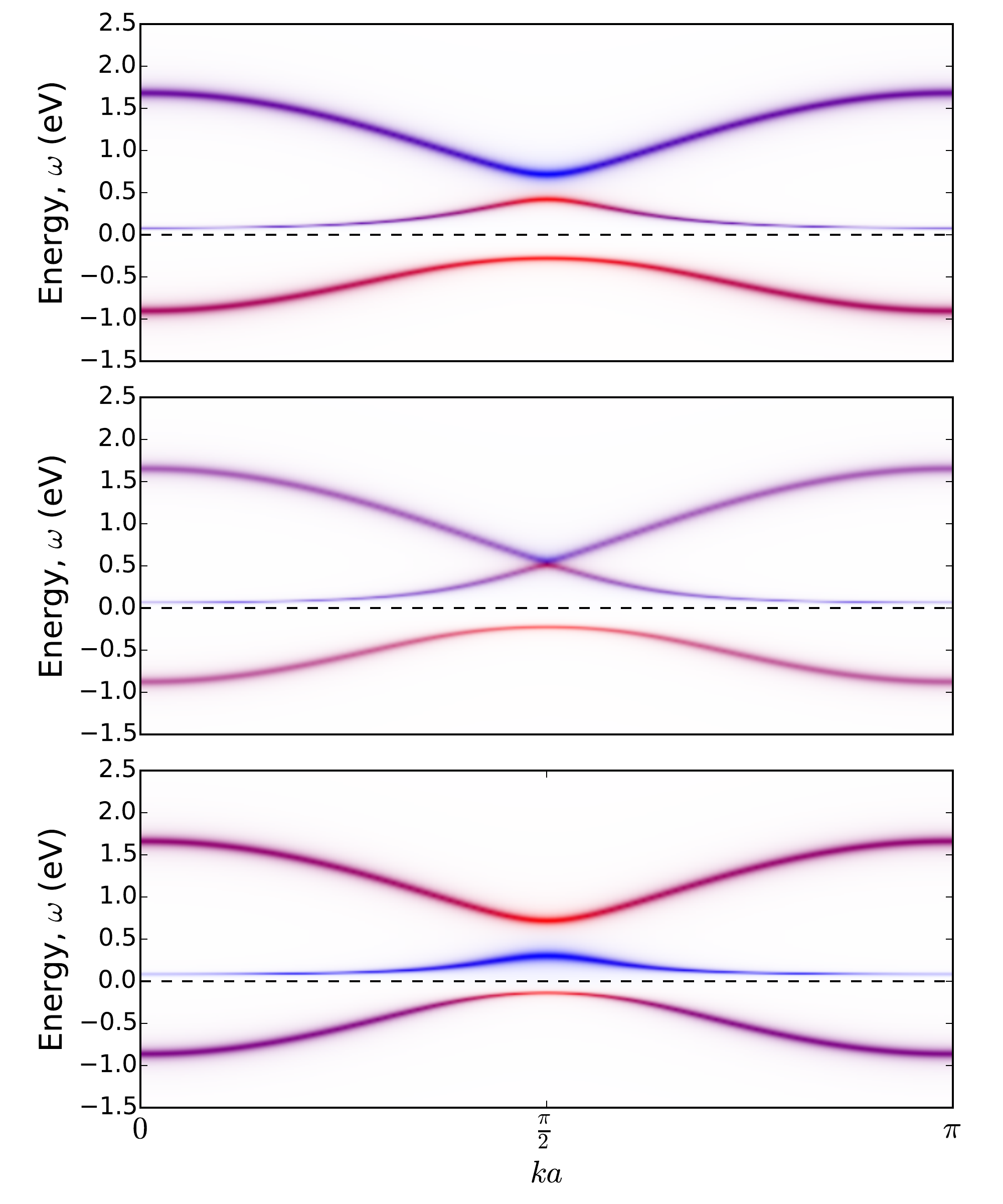}
\caption{\label{fig:RNO_opt_Fig12} $k$-resolved spectral function for the simple 1D model.
The parameters for the three panels are taken from the fitted parameters of the self-energies for three cases. 
Top panel: $J = 0.85$ eV, $U = 1.0$ eV, middle panel: $J = 0.8$ eV, $U = 1.2$ eV, bottom panel:
$J = 0.85$ eV, $U = 2.0$ eV. 
The color represents a site character of states: LB (red) and SB (blue).}
\end{figure}

Given that $\epsB > 0$, the first root $\omega_{-}$ is always the smallest, $\omega_{-} < \omega_{+}, \epsB$, and it corresponds to the occupied LB-like band. 
The order of the two other roots, on the other hand, depends on the parameter regime: 
\begin{enumerate}
\item
$\epsB \gg \epsA$ ($\Deltaeff_s\gg0 $), which implies $\epsB > \omega_{+}$: In this case an excitation bringing electrons from a LB site to a SB site is larger than a typical Mott excitation energy [between the corresponding lower (LHB) and upper (UHB) Hubbard bands] and the band right above the correlation gap (lowest unoccupied band, LUB) is predominantly of LB character.
The gap corresponds to a Mott gap on the LB sublattice, with the practically empty SB sub-lattice lying above the UHB, and in this limit one can talk about a ``site-selective Mott transition'' as introduced in Ref.~\onlinecite{Park2012}.
\item
$0<\epsB \ll \epsA$ ($\Deltaeff_s\ll 0$), which implies $\epsB < \omega_{+}$: The intersite hopping excitation LB~$\to$~SB in this case is smaller than the Mott excitation energy, the LUB is mainly of SB character, and this limit corresponds to a sort of an ``intersite charge-transfer'' (or ``Peierls-Mott'') regime.
One can say that in this regime the unoccupied SB-like band is situated between the LHB and UHB of the LB sublattice.
\end{enumerate}

\begin{table}[!h]\centering
\begin{tabular*}{\linewidth}{@{\extracolsep{\stretch{1}}}ccrrr@{}}\toprule
$U$ (eV) & $J$ (eV) & $\epsA$ (eV) & $\epsB$ (eV) & $\er^2$ (eV$^{2}$) \\
\midrule
$1.0$ & $0.85$ & 0.10 & 0.70 & 0.12 \\
$1.2$ & $0.80$ & 0.26 & 0.55 & 0.12 \\
$2.0$ & $0.85$ & 0.51 & 0.29 & 0.13 \\
\bottomrule
\end{tabular*}
\caption{Parameters of the simple model used to plot Fig.~\ref{fig:RNO_opt_Fig12}.
Note the crossover from one regime ($\epsA < \epsB$) to another $\epsA > \epsB$ as $U$ is increased. 
The correlation strength $\er$ is almost constant due to the choice of parameters $U$, $J$ along the line of the constant leading edge (see main text).}
\label{table:5}
\end{table}

These two regimes and the crossover between them (the crossover point is determined by a relation $\omega_{+} = \epsB$) can be nicely illustrated by the $k$-resolved spectral function evaluated according to Eq.~\eqref{eq:akw}, displayed in Fig.~\ref{fig:RNO_opt_Fig12}. 
The self-energy parameters are taken from the fits to the actual QMC data, where the first (dominant) pole of the double pole fit [Eq.~\eqref{eq:fit2pole}] is used for the LB self-energy (with $\er^{2} \equiv p_{1}$ and the small parameters
$p_{2}$, $p_{3}$ neglected).
The resulting values of parameters $\epsA$ and $\epsB$ are presented in Table~\ref{table:5}.

The lowest (occupied) band is predominantly of LB character reflecting the occupancy disproportionation ($n_{\LB} > n_{\SB}$) in the BDI phase.
The two regimes described above are easily distinguished by comparing the spectral functions for $U = 1.0$ (top panel in Fig.~\ref{fig:RNO_opt_Fig12}) and $U = 2.0$ (bottom panel). 
For $U = 1.0$ the character of the middle band (LUB) is predominantly LB-like (red), especially close to the Peierls-gap edge, and the top-most band is SB-like (blue). 
For $U = 2.0$ the characters of the two top bands are switched. 
An intermediate case corresponding to the crossover regime demonstrates a higly mixed character of these two bands.
\section{Results of GGA+DMFT calculations\label{Appendix:F}}
\begin{figure}
\includegraphics[width=\linewidth]{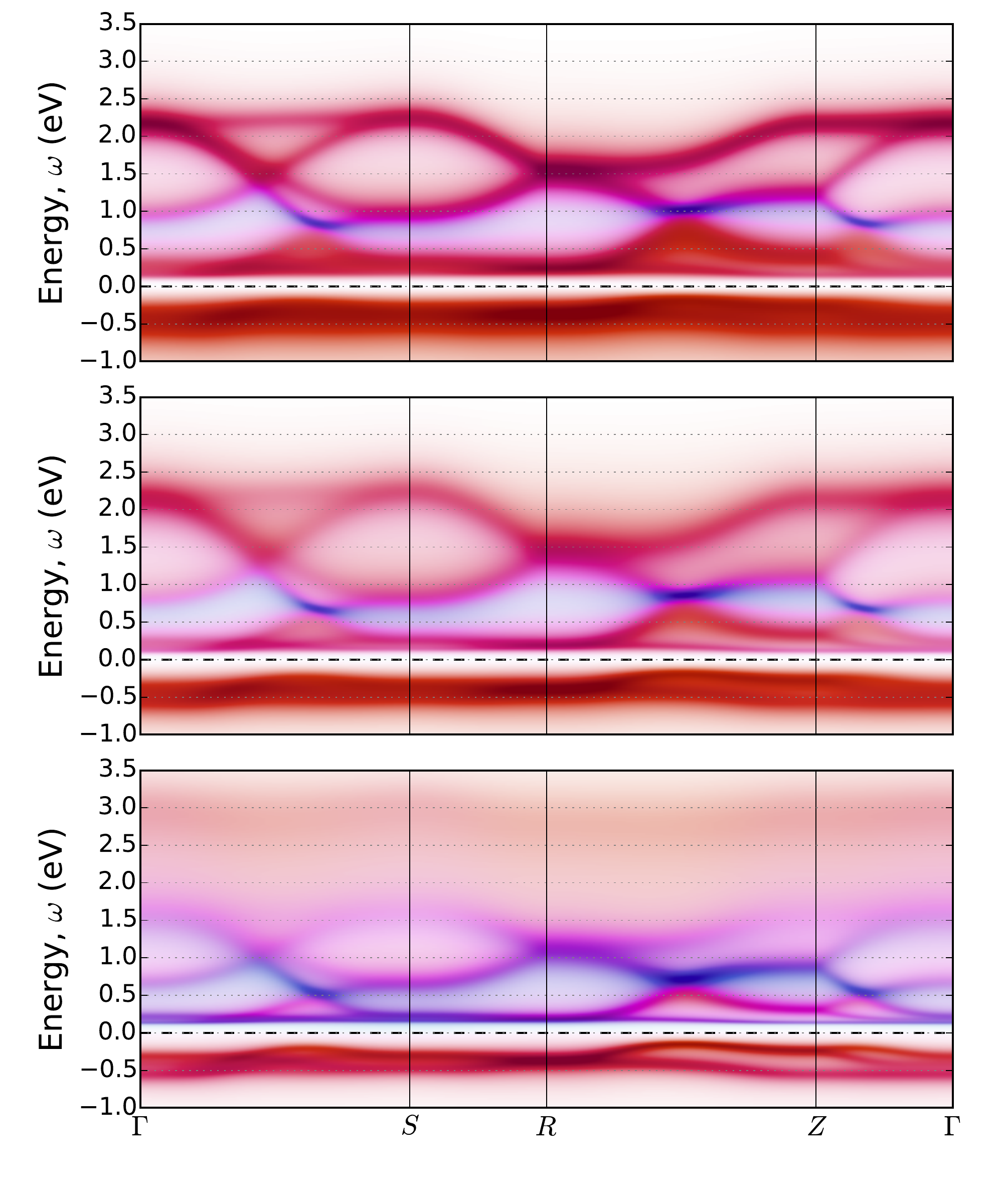}
\caption{\label{fig:RNO_opt_Fig13} $k$-resolved spectral function for SmNiO$_{3}$ as obtained in the GGA+DMFT calculation.
The parameters for the three panels are taken from the fitted parameters of the self-energis for three cases. 
Top panel: $J = 0.85$ eV, $U = 1.0$ eV, middle panel: $J = 0.8$ eV, $U = 1.2$ eV, bottom panel: $J = 0.85$ eV, $U = 2.0$ eV. 
The color represents a site character of states: LB (red) and SB (blue).}
\end{figure}

The above analysis provides a model picture of the electronic structure of the BDI phase at the qualitative level. 
It is not entirely clear to which of the two regimes (Mott or Mott-Peierls) the nickelates actually belong. 
The fact that the lower-energy peak in the infra-red spectrum has higher intensity may suggest that they are more in the Mott-Peierls regime, or in the crossover between the two regimes. 

Although the calculated GGA+DMFT electronic structure is more complex than the simple model above, many of the qualitative features can still be recognized. 

As one can see in Fig.~\ref{fig:RNO_opt_Fig13}, there is a clear difference in the dominant character of the lowest unoccupied states right above the gap.
Deep in the BDI phase ($U - 3J - \Delta_{s} \simeq -1.8$) these states are mainly of LB character (red) almost everywhere in the BZ. 
On the contrary, close to the upper boundary of the BDI phase ($U - 3J - \Delta_{s} \simeq -0.8$) the entire band above the gap is SB-like (blue). 
In the intermediate regime ($J = 0.8$, $U = 1.2$) the characters of the unoccupied states are mixed apart from  certain parts of $k$-space. 
It is also worth noting that an additional diffuse band appearing between 2.5 and 3.0 eV in the case of $U = 2.0$ is an atomic-like UHB that lies above the unoccupied SB bands and whose position relative to the occupied band (LHB) is given by $U + J$ (equal to 2.85 eV in this case).
This band is responsible for the transfer of a substantial amount of the spectral weight to higher energies in the one-electron spectra but plays no significant role in optical transitions because of its incoherent nature.

\end{document}